\documentclass[twocolumn,pra,epsfig]{revtex4}

\usepackage{epsfig}
\usepackage{amsmath}
\usepackage{dcolumn}
\usepackage{lscape,graphicx}
\usepackage{ifthen}
\newboolean{ShowNotes}

\setboolean{ShowNotes}{false} 

\def\bea{\begin{eqnarray}}
\def\eea{\end{eqnarray}}
\def\ben{\begin{equation}}
\def\een{\end{equation}}
\def\benu{\begin{enumerate}}
\def\enu{\end{enumerate}}

\def\n{n}

\def\sss{\scriptscriptstyle\rm}

\def\1var{(\bx_1...\bx\N)}

\def\half{\frac{1}{2}}

\def\br{{\bf r}}

\def\bx{{\br t}}

\def\bj{{\bf j}}

\def\x{_{\sss X}}
\def\c{_{\sss C}}
\def\s{_{\sss S}}
\def\xc{_{\sss XC}}

\def\N{_{\sss N}}
\def\H{_{\sss H}}

\def\ext{_{\rm ext}}

\def\sph_int{ {\int d^3 r}}

\input rotate

\def\tot{_{\rm tot}}
\def\bE{{\bf E}}
\def\bG{{\bf G}}
\def\bg{{\bf g}}
\def\bP{{\bf P}}
\def\R{_{\rm R}}
\def\L{_{\rm L}}
\def\F{_{\rm F}}
\def\eps{\epsilon}
\def\bGamma{{\mathbf \Gamma}}
\def\bSigma{{\mathbf \Sigma}}
\def\epsH{\epsilon_{\sss HOMO}}
\def\epsL{\epsilon_{\sss LUMO}}
\def\res{_{\sss res}}

\ifthenelse{\boolean{ShowNotes}}
{
  \newcommand{\note}[1]
  {\vspace{1mm}
   \noindent{\footnotesize\sffamily \(<\)Note\(>\) {\bf #1} \(<\)/Note\(>\)
\hfill}
   \vspace{1mm}}
}
{
  \newcommand{\note}[1]{}
}

\begin{document}


\title{Density functional calculations of nanoscale conductance}

\author{Max Koentopp}
\author{Connie Chang}
\affiliation{Department of Chemistry \& Chemical Biology, Rutgers University, 610 Taylor Rd, Piscataway, NJ 08854}

\author{Kieron Burke}
\affiliation{Department of Chemistry, UC Irvine, 1102 Natural Sciences 2, Irvine, CA 92697}

\author{Roberto Car}
\affiliation{Department of Chemistry and Princeton Institute for the
  Science and Technology of Materials (PRISM),
  Princeton University, Princeton, NJ 08544}

\date{\today}

\large

\begin{abstract}
Density functional calculations for
the electronic conductance of single molecules are now common.  We examine the 
methodology from a rigorous point of view, discussing where 
it can be expected to work, and where it should fail.  
When molecules are weakly coupled to leads, local and gradient-corrected
approximations fail, as the Kohn-Sham levels
are misaligned.   In the weak bias regime, XC corrections to the current
are missed by the standard methodology.
For finite bias, a new methodology for
performing calculations can be rigorously derived using
an extension of time-dependent current density functional theory from
the Schr\"odinger equation to a Master equation.
\end{abstract}
\maketitle
\tableofcontents

\newcommand{\upalda}{^{\scriptscriptstyle \rm ALDA}}
\newcommand{\downalda}{_{\scriptscriptstyle \rm ALDA}}
\newcommand{\vuc}{^{\scriptscriptstyle \rm VUC}}
\newcommand{\Tr}{{\rm Tr\; }}
\newcommand{\TrB}{{\rm Tr_{R}\; }}
\def\Trb#1{{\rm Tr\;}\left( #1 \right) }
\def\bE{{\bf E}}

\section{Introduction and Notation}
\label{s:intro}
Single molecules used as building blocks such as diodes, transistors, or
switches have attracted much interest as a basis for a future {\it molecular electronics}\cite{NR03}.  Many groups world-wide
have been performing either experiments or calculations.  
There has been tremendous progress, especially in the areas of metallic
wires\cite{ALR03,YRBA98,SJEU97,SACY98,TJ03} and nanotubes
\cite{ELHBGT96,TDDTSGD97,PTYGD01,N99,RYPGSM02,CR02,MP04}.  However, comparison between theory and
experiment has been much less successful for molecular electronics, i.e., organic molecules between
two electrodes.
Experimentally, obtaining
consistently reproducible results from device to device has been
problematic\cite{UEPV06,VKTN06}.  Theoretically, the challenge is finding a 
method to quantitatively determine device characteristics with neither 
empirical input nor
over-parameterization\cite{RCBH03,EWK04,KBE06,SZVD05}.

In recent years, density functional theory (DFT) calculations of electronic transport through
single molecules have been published by an ever increasing number of
research groups. We focus here on purely electronic transport calculated
with DFT-based methods. The most prominent method is 
the Landauer--type scattering formalism \cite{L57,L70,BILP85,B86},
formulated in terms of Greens functions in combination with ground state
DFT. It can be derived using the Keldysh non-equilibrium Greens-function
formalism \cite{MW92,BMOT02}. In the following, we will call this method
{\em the standard approach}. It can also be obtained from elementary scattering
theory (e.g.\ \cite{XDR02,EWK04}), or using the Kubo linear
response formula\cite{FL81,BS89,KK01}.  However, such derivations are limited\
to at best Hartree-interacting electrons, as we discuss. 
Apart from calculating the
current--voltage characteristics of a coherent molecular junction in the
Landauer scattering picture
\cite{EWK03,EWK04,DPLb00,DPL00,DPL02,BMOT02,EK01,XDR02,DGD01,KBY04,BN03,P05b,TJ05}, several
 additional aspects such as electron--phonon
coupling\cite{SRG06,FBLJ04,CZD04}, conformation--induced
switching\cite{JZCN06,ZDCZ04,DMG04,KBW02,GRD04},
or interaction with light\cite{ZDCZ04} have also been addressed
within these methods. Questions discussed also include the
influence of electrode-molecule 
bond geometry\cite{EWK03,EWK04,XR03,M06,BR03,HZLN05}
or effects of a gate electrode\cite{KBE06,KBY05,GRD04,BK03}.

Whereas ground state DFT has become
quite reliable for calculating the electronic structure and other
properties of molecules and solids\cite{FNM06}, this success has not extended to
transport calculations 
through organic molecules\cite{EWK04,KBE06}.
While these calculations were originally greeted with much enthusiasm,
researchers in the many-body community have always been skeptical\cite{RWHS06,HWWS03,HS02}.
A simple question to ask is, can such calculations correctly describe
the Coulomb blockade regime?  That the answer is patently no has
raised doubts about the validity of this approach among that community.
These doubts have been further compounded by the fact that
calculations within this
scattering framework usually overestimate the experimentally measured
conductance of organic molecules by about one order of magnitude \cite{EWK04,KBE06}.
Only for transport through atomic metallic wires\cite{ALR03,YRBA98,SJEU97,SACY98},
do calculations yield results in agreement with experiment, but this
is not a true test of the method, as the same result is found in any
calculation yielding a unit of conductance per channel.

The performance of this standard approach is the main subject of
this review.  As we describe below, neither of the traditional
tests of DFT calculations, i.e., direct comparison with experiment
or benchmark testing against more accurate theoretical methods for
smaller systems, are generally available for this problem.
On the one hand, the experimental situation is often not well-characterized,
while on the other, these are transport calculations with systems
of up to several hundred atoms.  Alternative treatments are either prohibitively
expensive or of such a modelistic nature as to not allow meaningful
tests.  Thus, at present, there is no simple way to know when the standard
approach is accurate or reliable.  Instead, we examine carefully DFT treatments, and show that
the standard approach is an approximation to a more general approach
using time-dependent DFT, and from this perspective, its limitations
can be deduced.

Ground-state DFT
is based on rigorous theorems and so, if correctly applied to a problem, using 
a sufficiently accurate approximate functional, will produce an accurate result.
The purpose of the present article is to ask two simple questions:
(a) is the present {\it standard approach} formulation derivable from the basic theorems 
of ground-state DFT, and
(b) if so, are our present approximations sufficiently accurate for
conductance calculations?  The answers show a variety of deficiencies
(e.g.\ inadequacy of the ground-state approximation, 
approximations made by using local functionals)
in the present theory and we do not yet know
how important these drawbacks are.  We don't know how frequently situations
are encountered in which these limitations are quantitatively significant.

We discuss here three major issues that need to be resolved to improve on the present
state of transport calculations. (i) The first involves the
accuracy of ground-state functionals.  The functionals presently used in the Landauer and Kohn-Sham
approximations might not capture enough
of the physics to be useful and more importantly they might not give good
qualitative or quantitatively accurate results.  The worst defect we have found
is due to lack of derivative discontinuity in LDA and GGA functionals, which
leads to an artificial level broadening and can greatly overestimate the
conductance~\cite{KBE06}, due to incorrect positions of the resonances if the molecule is weakly coupled to the leads\cite{TFSB05,KBY06}.
(ii) The second issue is the missing exchange-correlation (XC) contribution in the Landauer formula.
Present calculations  entirely miss this contribution to the current.  Some groups have sought to improve 
on the issues delineated in (i) and (ii) by calculating the XC corrections to the current via the gradient expansion
corrections in the Vignale-Kohn
approximation \cite{SZVD05} and the exact exchange Kohn-Sham potential with the Optimized
Effective Potential (OEP) \cite{KBY06}.  The exact exchange potential can also be estimated
with self-interaction corrections (SIC) \cite{TFSB05, TS07} where the self-interaction errors in LDA
DFT calculations are subtracted out.  Furthermore, in the weak bias limit, a careful
application of the Kubo
response equation coupled with the DFT formulation, can be used to find the
exact answer \cite{KBE06}.  (iii) The third issue we address is an exact
theory for finite bias, since transport experiments are often conducted under
a finite voltage drop.  Thus, an exact formula couched in DFT terms must be
derived for these conditions.

In the last few years, the DFT computational transport community
has become aware of these issues\cite{EWK04,KBE06,SZVD05,TFSB05}, and a variety of approaches
to overcome them have been suggested.   Many are looking to alternative
formulations, such as configuration-interaction (CI)
in quantum chemistry\cite{DG04} or GW in many-body physics\cite{PDGN05,TR07},
to include effects that are missed in present ({\em standard approach}) DFT treatments.  Such
calculations are sorely needed, to test the DFT formulations against and learn
their limitations.  Accurate wavefunction treatments are extensively used in both ground--state and 
TDDFT as benchmarks and to provide insight into functional development\cite{FBLB02}.
But since such calculations are typically far more expensive than DFT
calculations, and given the chemical complexity of the devices that
can be built, there remains an over-riding need to develop a reliable
DFT approach.

Thus a variety of new DFT-based formulations of the problem are being developed.
One discussed here includes
using a Kohn-Sham effective single
particle version of a Master equation formulation of transport\cite{BCG05}
which will be discussed in section \ref{s:mastertheory}. Using TDDFT,
another approach obtains the current by calculating the time evolution of
a system consisting of a molecule coupled to two
finite metallic contacts and turning on  a potential step, resulting
in two different chemical potentials\cite{SA04,SAb04}.
A third is to use large finite leads, and watch a capacitance discharge\cite{VT04}.
All three methods essentially begin from a static distribution,
apply some change, and allow the system to evolve to a steady,
but non-equilibrium distribution.  For non-interacting electrons
in the weak bias limit, all agree, both with each other and
the standard approach, but likely disagree in the
general case of interacting electrons in finite bias.
Under certain limiting conditions, such as adiabatic approximations
to TDDFT, and local approximations to ground-state DFT,
they yield the same results.

Because of the breadth of topics we cover in this review, we have
collected the notation used in various formulas and expressions in
Table~\ref{t:notation} for easy reference.  We use atomic units
($e^2=\hbar=m=1$) throughout, unless otherwise stated. So all energies
are in Hartree (1H~=~27.2eV~=~627kcal/mol) and 
distances in Bohr-radii (0.529\AA).
\begin{table}
\caption{\label{t:notation} Notation for formulas}
\begin{ruledtabular}
\begin{tabular}{ll}
  \multicolumn{1}{c}{symbol} & \multicolumn{1}{c}{description}\\\hline  
 $n(\br)$ &  electron density as a function of position\\
 $v\ext(\br)$ & ext. potential due to nuclei and ext. fields\\
 $v\xc(\br)$ & exchange-correlation potential\\
 $v\H(\br)$ & Hartree potential \\
 $v_{tot}(\br)$ & $v\ext(\br)+v\H$(\br), total electrostatic potential\\
 $v\s(\br)$ & KS potential$=v\ext(\br)+v\H(\br)+v\xc(\br)$\\
 $\chi\s(\br,\br',\omega)$ & KS density-density response function\\
 $\chi_{prop}(\br,\br',\omega)$& proper susceptibility\\
 & (density-density response function)\\
 $\chi(\br,\br',\omega)$ & many-body density-density response function\\
 $i$ & occupied level index\\
 $a$ & unoccupied level index\\
 $q$ & transition index of transition $i\rightarrow a$\\
 $\Phi_{q}(\br)$ & $\Phi_i^*(\br)\Phi_a(\br)=$ KS ground state orbitals\\
 $\omega_q$ & $\epsilon_a-\epsilon_i$\\
 $\alpha=L/R$ & quantum numbers for (left/right) lead \cr
              & electrons\\
 $i,j$ & indices for KS orbitals of the device region \\
 $\alpha,\beta$ & cartesian indices\\
 $\epsilon_{k\alpha}$ & energy of electron in lead 
$\alpha=L/R$\cr
& with momentum $k$\\
$V_{k\alpha,n}$ & coupling between leads and molecule\\
$c^{\dagger}_{k\alpha}(c_{k\alpha})$ & creation (destruction) operator for 
  electron \cr & with momentum $k$ in lead $\alpha$\\
$d^{\dagger}_{n}(d_{n})$ & creation (destruction) operator for 
  electron \cr & with quantum numbers $n$ on the molecule\\
$\rho_{\alpha}(\epsilon)$ & density of states in lead $\alpha$\\
$f_{l/R}(\epsilon)$ & Fermi-Dirac distribution functions in leads\\
$\Gamma^{L/R}_{i,j}$ & transition rate to left(right) lead\cr
&$=2\pi\sum_{\alpha\epsilon L/R}\rho_{\alpha}(\epsilon)V_{\alpha,i}(\epsilon)
V_{\alpha,j}^{*}(\epsilon)$\\
$\bg^{r}$ & surface Green's function for the leads\\
${\mathbf \tau}$ & hopping matrix describing the coupling \cr & between
 leads and molecule,\\
 & its elements are given by $V_{k\alpha,n}$\\
${\mathbf \sum}_{R}({\mathbf \sum}_{L})$& self-energy matrices $={\mathbf \tau}
 \bg^{r}{\mathbf \tau}^{\dagger}$\\
$\bG^{a(r, >, <)}$ & full advanced (retarded, greater, lesser)\cr 
&  Green's function for the extended molecule\\
$\bg_{\sss 0}$ & unperturbed KS Green's function for device\\
$\Psi\s$ & TDDFT KS wavefunction of the entire system\\
$\Psi_{\alpha}$ & TDDFT KS wavefunction projected on the\\
&  leads $\alpha=L,R$\\
$\Psi_{C}$ & TDDFT KS wavefunction projected on the \\
& central (molecule) region\\
$H_{\alpha \beta}$ & block of the TDDFT KS Hamiltonian with\\
& $\alpha,\beta=$ L(left), R(right), C(center/molecule)\\ 
$\epsilon_{res}$ & position of level in a resonant tunneling device\\
$\gamma$ & width of resonance in a resonant tunneling \cr & device/  
coupling to the leads\\
$T(\epsilon)$ & transmission coefficient as a function of energy\\
$n(\epsilon)$ & spectral density of states\\
$f$ & occupation of level\\
$\epsilon_F$ & Fermi energy\\
$\epsilon_{HOMO}(\epsilon_{LUMO})$ & HOMO (LUMO) level of device\\
$\sigma(\br,\bf{r'},\omega)$ & conductivity (current-current response \cr 
& function)\\
$S_{T} $ & density matrix for total system \\
$S$ & reduced density matrix
\end{tabular}
\end{ruledtabular}
\end{table}

\section{Review of the standard approach}
\label{s:revtrans}
\label{s:trans}
In this section, we first review
standard calculations that utilize a combination of ground--state DFT
and the Landauer scattering formulation (the {\em standard approach}). We then look at approximations commonly employed in the course of calculating the conductance with this method.  
\subsection{Landauer scattering formulation}
\label{s:Landauer}

\begin{figure}[htb]
\unitlength1cm
\begin{picture}(10.8,4.1)
\put(-6.2,-4.8){\makebox(12,9){
\includegraphics{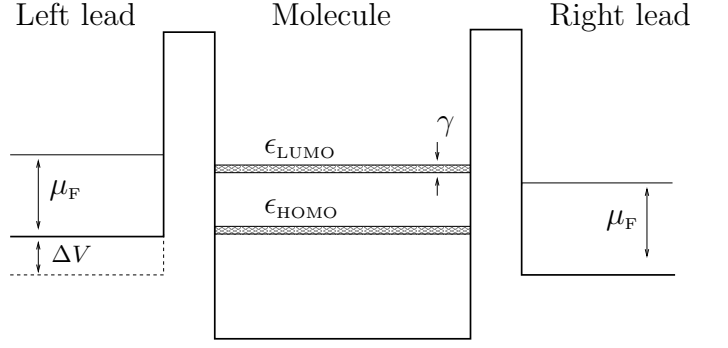}
}}
\put(3.3,3.9){\large Molecule}
\put(-0.1,3.9){\large Left lead}
\put(7.0,3.9){\large Right lead}
\put(7.75,1.25){\large $\mu_{\sss{F}}$}
\put(3.2,2.2){\large $\epsilon_{\sss{LUMO}}$}
\put(3.2,1.4){\large $\epsilon_{\sss{HOMO}}$}
\put(5.5,2.5){\large $\gamma$}
\put(0.35,1.65){\large $\mu_{\sss{F}}$}
\put(0.27,0.75){ $\Delta V$}
\end{picture}
\caption{Schematic of the potential of 
a resonant tunneling device with the LUMO level of
 the device molecule sandwiched between the Fermi levels of the leads which are shifted relative to each other by an applied bias voltage $\Delta V$.  When $\Delta V=0$, the chemical potential is $\mu_F$.  The LUMO and HOMO levels have a small width $\gamma$, indicating weak coupling to the leads.} 
\label{f:barrier}
\end{figure} 
The Landauer scattering formulation can be easily understood in terms
of a simple schematic for a molecular tunneling device, as shown in
Fig.~\ref{f:barrier}.  In this cartoon, the electrons are
non-interacting and the system is one
dimensional. The leads are featureless boxes, and the `molecule' consists
of states localized to the barrier region.  In the cartoon, $\mu_F$ is
the chemical potential of the entire system, i.e., in its ground state and in the absence of
a bias.  The molecule has been drawn so as to be weakly coupled to
the leads, so that the levels on the molecule are only slightly broadened
into resonances of width $\gamma$.

In the Landauer picture, the applied bias raises the potential on the
left lead by an amount $\Delta V$.  There is now an imbalance in the
system.  If one waits a long enough time,
eventually many electrons would flow from the left to the right, and
re-establish equilibrium, with a common chemical potential.  This is
not the regime we are interested in.  Instead, on an intermediate time-scale, one assumes a
steady current is established, that is sufficiently small as to have
no effect on the reservoir levels.

The current can then be calculated from the two-terminal Landauer formula\cite{L57}
which, for this case, is simply
\ben
\label{e:basic_landauer}
I=\frac{2}{\pi}\int^{\infty}_{-\infty}d\epsilon\Delta f \, T(\epsilon)
\een
where
\bea 
\Delta f&=&f\L(\epsilon)-f\R(\epsilon)\cr
&=&f(\epsilon-\mu\F-\Delta V)-
f(\epsilon-\mu\F).
\eea
and $T(\epsilon)$ is the transmission coefficient at energy $\epsilon$.  The 
factor of two is for the two spin channels.

\begin{figure}[htb]
\begin{center}
\includegraphics[angle=0,width=6cm]{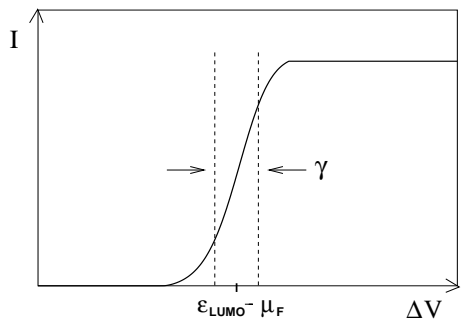}
\end{center}
\caption{Schematic current-voltage characteristic of the  
 resonant tunneling device displayed in Fig.~\ref{f:barrier}.
The onset of the current occurs around $\epsilon{\sss LUMO} - \mu\F$. The step
is broadened by the coupling $\gamma$.} 
\label{f:iv_schematic}
\end{figure}

This can be easily understood as follows.  Considered as a function of
energy, only those states in the window between $\mu_F$ and $\mu_F+\Delta V$
can carry a net current.  Those below are occupied on both sides, those
above are unoccupied on both sides.  Each state of energy $\epsilon$ in the window will transmit
an electron with probability $T(\epsilon)$, yielding that contribution
to the net current.  The schematic shown in Fig.~\ref{f:barrier} will yield a current-bias
curve like that shown in Fig.~\ref{f:iv_schematic}
The differential conductance, $dI/d\Delta V$,
will be strongly peaked in the position of the LUMO.
The result is even simpler in the zero bias limit, as $dV \to
0$. Then, $df = \delta (\mu\F)dV$, so that the
conductance becomes
\ben
G = \frac{dI}{dV} = \frac{2T (\mu\F)}{\pi}.
\een

Thus for non-interacting electrons in a fixed potential, the Landauer
formula is easily understood and justified. 

\subsection{Interacting Electrons}
\label{s:MW}

The Landauer formula for non-interacting (or at most
Hartree-interacting) electrons was later generalized to interacting
electrons by Meir and Wingreen \cite{MW92}, who 
formulated an algorithm for calculating the current 
using the full non-equilibrium Green's functions for the system.  They
employ a second quantized Hamiltonian description for the electrons in the
leads, the interacting region (molecule), and the coupling between them.  Initially
uncoupled, the coupling
between the leads and the molecule is turned on slowly via the $V_{ka,n}$
term in Eq. (\ref{e:Ham}):
\bea
\label{e:Ham}
H&=&\sum_{k,a\,\epsilon\, L,R}
\epsilon_{ka}c^{\dagger}_{ka}c_{ka}+H_{int}(\{d^{\dagger}_{n}\};\{d_{n}\})\cr
&+&\sum_{k,a\,\epsilon\,
  L,R}(V_{ka,n}c^{\dagger}_{ka}d_n+H.c.).
\eea  
Here, $k$ refers to the momentum of an electron with energy $\epsilon_{ka}$ in the left or right lead, labeled by $\alpha$. The creation and annihilation operators are denoted by $c^{\dagger} (c)$ and $d^{\dagger} (d)$, referring to the leads and the molecule, respectively.
Then, using the continuity equation for the current, the Keldysh formalism for the
Green's functions and allowing the electrons in the device region to
interact while keeping the electrons in the leads noninteracting, they
find an expression for the current when the leads are at different chemical
potentials: 
\bea
\label{e:MW}
I&=&{{2}\over{\pi}}\int{d\epsilon(tr\{[f_{L}(\epsilon})\Gamma_{i,j}^{L}-f_{R}(\epsilon)
\Gamma^{R}_{i,j}](G_{i,j}^{r}-G_{i,j}^{a})\}\cr
&+&tr\{(\Gamma^{L}_{i,j}-\Gamma^{R}_{i,j})G_{i,j}^{<}\}),
\eea
and
$\Gamma_{i,j}^{L/R}=2\pi\sum_{a\,\epsilon\,L/R}\rho_{a}(\epsilon)V_{a,i}(\epsilon)V_{a,j}^{*}(\epsilon)$
where $i,j$ indexes the states in the interacting region and $a$ indexes the
states in the leads.  $\bG^r$, $\bG^a$, and $\bG^<$ refer respectively to the retarded, advanced, and lesser Green's functions. 
Meir and Wingreen\cite{MW92} derived a simpler formula for the case of
proportional couplings ($\Gamma^{R}_{i,j}=\alpha\Gamma^{L}_{i,j}$) 
\ben
\label{e:current_prop}
I={2\over{\pi}}\int{d\epsilon}\left[f_{\sss L}(\epsilon)-
f_{\sss R}(\epsilon)\right]Im\left[tr\left\{\bGamma \bG^r\right\}\right],
\een
where $\bGamma={{\bGamma^{\sss L}\bGamma^{\sss R}}/{(\bGamma^{\sss L}+\bGamma^{\sss R})}}$.
 This however, is a strong limitation due to  their reliance on
symmetric contacts which is never fulfilled for a realistic system.
Only an atomic point contact could satisfy this condition since it
requires that each orbital on the device couples equally to the left and right
contact.
This restriction can be removed\cite{EWK04,SA04}, resulting in
a general formula for the current in terms of the non-equilibrium, 
self-consistent Green's function.

\subsection{The {\em standard approach}}
\label{s:DFT-NEGF}
\begin{figure}[tbh]
\begin{center}
\leavevmode
\includegraphics[angle=0,width=7cm]{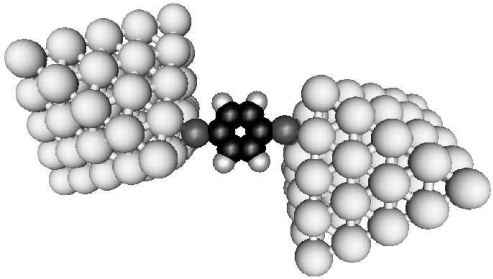}
\end{center}
\caption{Landauer approach: schematic representation of a
  benzene-1,4-di-thiol molecule between two gold contacts. The
  molecule plus gold pyramids (55 atoms each) constitute the {\em extended
  molecule} as used in the DFT calculations for the {\em standard approach}.}
\label{f:fig1}
\label{f:benzene}
\end{figure}

Because of the difficulties involved in solving the full many-body 
problem for the non-equilibrium Green's
functions exactly for an interacting system of many electrons,
 the Green's function in Eq.~(\ref{e:current_prop}) is usually approximated with the ground--state Kohn-Sham effective single-particle Green's
 function (DFT-NEGF).
This complicates the simple picture of Fig.~\ref{f:barrier} somewhat, as the KS
potential changes with the applied bias.  A self-consistent KS
potential must be found, which will continuously change from
being raised by $\Delta V$ on the left, to being at its equilibrium
level on the right.  These changes will not be confined solely to the
molecule, but should die off within one or two Fermi wavelengths into the
leads.  Thus one must define an extended molecule as in Fig.~\ref{f:benzene}, which
includes those parts of the leads where the KS potential differs from
its non-biased value.

Also note that in most calculations, the molecule is chemically bonded
to the leads.  Thus its levels will be much broader than pictured in
Fig.~\ref{f:barrier}, overlapping with one another, and delocalizing into the leads.

In addition, approximations to the self energy matrix
 $\bSigma_R$ and $ \bSigma_L \unboldmath$ are made (e.g.\ \cite{XDR02,DPL00,EWK03}).
 The  coupling between the right lead and the device is described by the hopping
 matrix ${\mathbf \tau}_R$ (whose elements are just the coupling terms $V_{ka,n}$ in
 Eq. (\ref{e:Ham})) and similarly for the coupling between the left lead
and the device.  There is no direct coupling
between the leads as this would cause electrons from the left lead to
 run into the right lead until a global equilibrium was reached.
 Then, the self-energy that encapsulates the effects
of coupling the left contact to the device can be written as:
\ben
\label{e:selfenergy}
\bSigma_L={\mathbf \tau}_{\sss L} \bg^r_{\sss L}{\mathbf \tau}_{\sss L}^{\dagger}
\een 
where $\bg^{R}_{\sss L}$ is the surface Green's function for the left lead.
An equivalent expression can be derived for the right contact.  The full 
Green's function for the device region,$\bG$, can then be written in terms of
the unperturbed KS Green's function, $\bg^{r}_0$,  for the extended
molecule (molecule plus small parts of the leads, Fig.~\ref{f:benzene}) as
\ben
\bG^{-1} =  \bg_0^{-1} + \bSigma_L +\bSigma_R.
\een
The coupling matrices $\bGamma^{R(L)}$ are given in terms of the
self-energy matrices as
\ben
\bGamma^{L(R)}=-i(\bSigma_{L(R)}-\bSigma^{\dagger}_{L(R)})=2\Im\bSigma_{L(R)}.
\een
 
 This 
self-energy only describes hopping onto and off the device from the leads, but
neglects the other processes that can occur in the leads.  The leads are thus
assumed to be non-interacting.  In this situation, a Dyson equation for the device region leads to the formula for the current for the case of 
non-interacting electrons derived from the more general expression given in Eq.~(\ref{e:MW}):
\ben
I={2\over{\pi}}\int{d\epsilon}\left[f_{\sss L}(\epsilon)-
f_{\sss R}(\epsilon)\right]tr\left\{\bG^a\bGamma^{\sss R}\bG^r \bGamma^{\sss L}\right\}.
\een
Identifying the trace with a transmission coefficient, this is
identical to Eq.~(\ref{e:basic_landauer}).
This approach has been implemented successfully by many groups by now
(e.g.\
\cite{DPLb00,DPL00,EK01,EWK03,AWE07,BMOT02,XDR02,DPL02,STB03,KBY04,KBY05b,RS04,BN03,P05b,TJ05}), and calculations have been performed for a large variety of  molecules, coupled to electrodes of different materials like gold, platinum and silicon. Molecules studied vary from $H_2$ to
alkyl-chains, to molecules built from aromatic subunits, or metallo-organic complexes.
 With sucessful we refer to a calculation in
 which the described standard formalism has been implemented and
    applied in a correct way. Especially early DFT based calculations of
    molecular conductance often had some shortcomings in the
    implementation or application, e.g. the coupling to the
    macroscopic leads was not accounted for in an
    appropriate way, which lead to strong deviations in the calculated
    current. Sucessful implementations, when applied to the same molecular system
    and contact geometry with the same functional performed by different groups yield similar results. 
However, the calculated conductances usually overestimate the experimentally measured
ones by about one order of magnitude \cite{EWK04,KBE06}.
By far the best studied molecule is benzene-dithiol coupled to gold
contacts via the sulfur
atoms\cite{EWK04,DPL00,BK03,STB03,XR03,FLS05,SRH04, KBY05b, KBY05c,TJ05,GKB06,GSW06,KKT06,SGP06}. Fig.\ref{f:DFT_HF} shows
results from a calculation\cite{KBE06} using the geometry in Fig.~\ref{f:benzene}
and comparing DFT with HF. Other calculations use slightly different
geometries, like e.g.\ planar gold surfaces and a super cell. Using
the latter geometry, calculations\cite{PGB07} employing the Master Equation approach yield a zero bias conductance of $\approx 0.4 G_0$, in very good agreement with Transiesta
calculations\cite{STB03} ($\approx 0.4  G_0$) and the results in Fig.~\ref{f:DFT_HF}
(0.3~$G_0$), which both employed slightly different geometries.  

\subsection{Limitations of the {\em standard approach}}
\label{s:lims}

Transport is inherently a time-dependent non-equilibrium problem with current
flow which is not at all within the domain of validity of a ground-state DFT
description. Its natural description is found within time-dependent
(current) DFT.  
Therefore, using ground state DFT to calculate the device Green's
function incorporates several uncontrolled approximations and errors
which will be investigated in sections \ref{s:inad} and \ref{s:weak}).
The time-dependent (TD) XC potential can be very different from its ground--state counterpart.  First, the step from time-dependent DFT to ground state DFT misses all
dynamic effects, i.e., the adiabatic approximation of section \ref{s:tddftapprox}. Also, for example, the partitioned system is assumed
to be in equilibrium and disconnected from the leads at some initial time.
Slowly, a finite voltage is turned on which shifts the chemical potentials of
the leads.  This necessarily drives the system out of equilibrium and so the
electron distribution in the leads do not follow the equilibrium Fermi
distributions. This effect is probably small, but its presence points to the
many problems with this approach.

Other problems with this approach include the incorrect placement of energy
levels due to the missing derivative discontinuity when a 
local approximation to the XC-functional is employed.  This leads to an error in the location of the
resonance peaks (section \ref{s:inad}).  The failure of local functionals to reproduce the
derivative discontinuity also produces resonance peaks that are too wide in energy which
results in a general  overestimation of the
conductance.

\section{Inadequacy of ground-state approximations}
\label{s:inad}

In this section, we take the {\em standard approach} prescription at face value, 
and assume it would give the correct conductance if implemented with
the exact ground-state density functional.  (In the next sections, we
will show that this is unlikely to be true in general.)  But we ask the simple question:
using present standard density functional approximations (LDA, GGA,
hybrids), will we get accurate results?  To answer this question,
we must first review some facts that are well-known in the DFT community.

\subsection{Exact ground-state DFT}
\label{s:gsDFT}

\begin{figure}[htb]
\unitlength1cm
\begin{picture}(12,9.6)
\put(-5,-4.6){\makebox(12,9){
\includegraphics{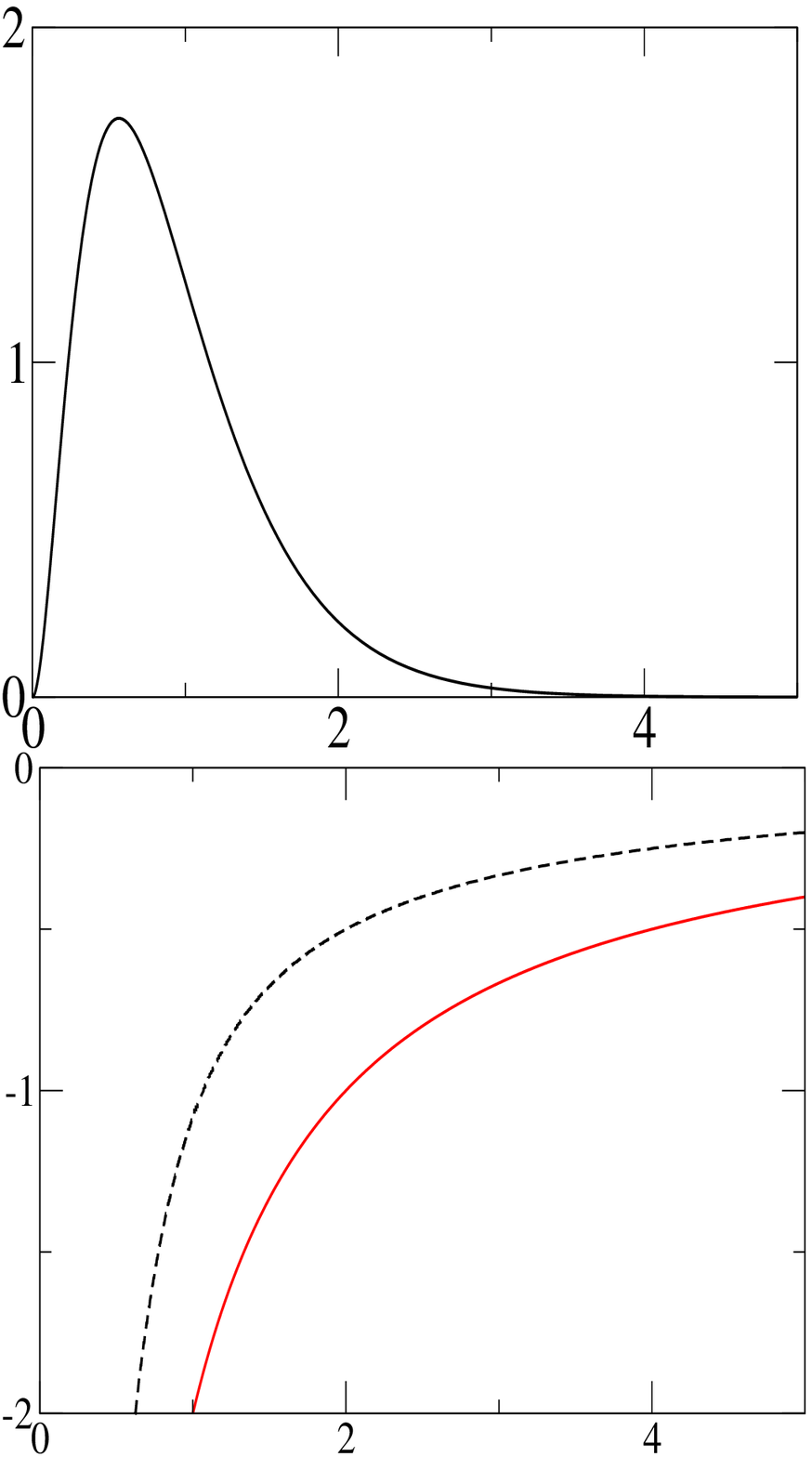}
}}
\setbox6=\hbox{\Large $4\pi r^2 \n(r)$}
\put(0.6,6.2) {\makebox(0,0){\rotl 6}}
\setbox6=\hbox{\Large $v(r)$}
\put(0.65,2.3) {\makebox(0,0){\rotl 6}}
\put(3.5,-.2){\Large $r$}
\put(3.8,1.8){\Large $-Z/r$}
\put(2.6,2.9){\Large $KS$}
\end{picture}
\caption{Top panel -- exact radial density for the He atom.
  Bottom panel -- external and exact KS potentials for the He atom (atomic units).}
\label{f:Hedens}
\end{figure}

In the top panel of Fig.~\ref{f:Hedens} we show the exact density $\n(\br)$ of the He atom, calculated by
Umrigar et al.\cite{UG93}, using quantum Monte Carlo to minimize the energy of
a highly accurate wavefunction.   In the bottom panel of Fig.~\ref{f:Hedens} we plot both the nuclear (i.e.\ external potential) potential, which is $-Z/r$ in atomic units, and the exact Kohn-Sham
potential $v\s(\br)$ for this system.  Two non-interacting electrons, inserted
in this potential, reproduce {\em exactly} the density of
the top panel of Fig.~\ref{f:Hedens}.
By the Hohenberg-Kohn theorem\cite{HK64}, this potential (if it exists) is unique.
All modern Kohn-Sham density functional calculations\cite{KS65} are calculations
of these fictitious non-interacting electrons in a KS potential.  The
goal of much research in DFT is to provide ever
more reliable approximations to the exchange-correlation (XC) energy, $E\xc[\n]$. Its
functional derivative with respect to density provides
the Kohn-Sham potential via
\ben
v\s(\br)=v\ext(\br) + v\H(\br) + v\xc(\br),
\label{e:vs}
\een
where $v\ext(\br)$ is the original external potential, $v\H(\br)$ is
the Hartree (or classical or electrostatic) potential, and 
$v\xc[\n](\br) = \delta E\xc[\n]/\delta \n(\br)$ is the XC
potential.
In traditional ground-state DFT, the
eigenvalues of the KS potential, $\epsilon_i$,
have no formal meaning, except that the
eigenvalue of the
highest occupied molecular orbital (HOMO) is exactly the negative of
the ionization energy (Koopman's theorem), as can be shown by studying the asymptotic decay
of the density\cite{PPLB82}.
However, the lowest unoccupied molecular orbital (LUMO) eigenvalue does not in general match the
negative of the electron affinity, i.e.,
\ben
\epsilon_{\sss HOMO}=-I,~~~~\epsilon_{\sss LUMO}\neq-A.
\een

In He, $\epsilon_{\sss 1s}=\epsilon_{\sss HOMO}=-0.903\,H$, but
$\epsilon{\sss 2s}=\epsilon_{\sss LUMO}=-0.158\,H $, while the electron
affinity $A=0$.
Thus the fundamental gap, $I-A$, is not equal to the
KS gap, i.e, the HOMO-LUMO energy difference, see Table~\ref{t:heKS}. 

\begin{table}
\caption{\label{t:heKS} Exact Kohn-Sham energies for the He atom,
  i.e.\ the orbitals for the KS potential of Fig.~\ref{f:Hedens}.}
\begin{ruledtabular}
\begin{tabular}{cd}
  \multicolumn{1}{c}{orbital} & \multicolumn{1}{c}{energy [H]}\\\hline  
 1s &  -0.90372436\\
 2s & -0.15773164 \\
 2p & -0.12656995 \\
 3s & -0.06454705

\end{tabular}
\end{ruledtabular}
\end{table}

\begin{figure}[htb]
\unitlength1cm
\begin{picture}(10.8,7.1)
\put(-5.8,-4.8){\makebox(12,9){
\includegraphics{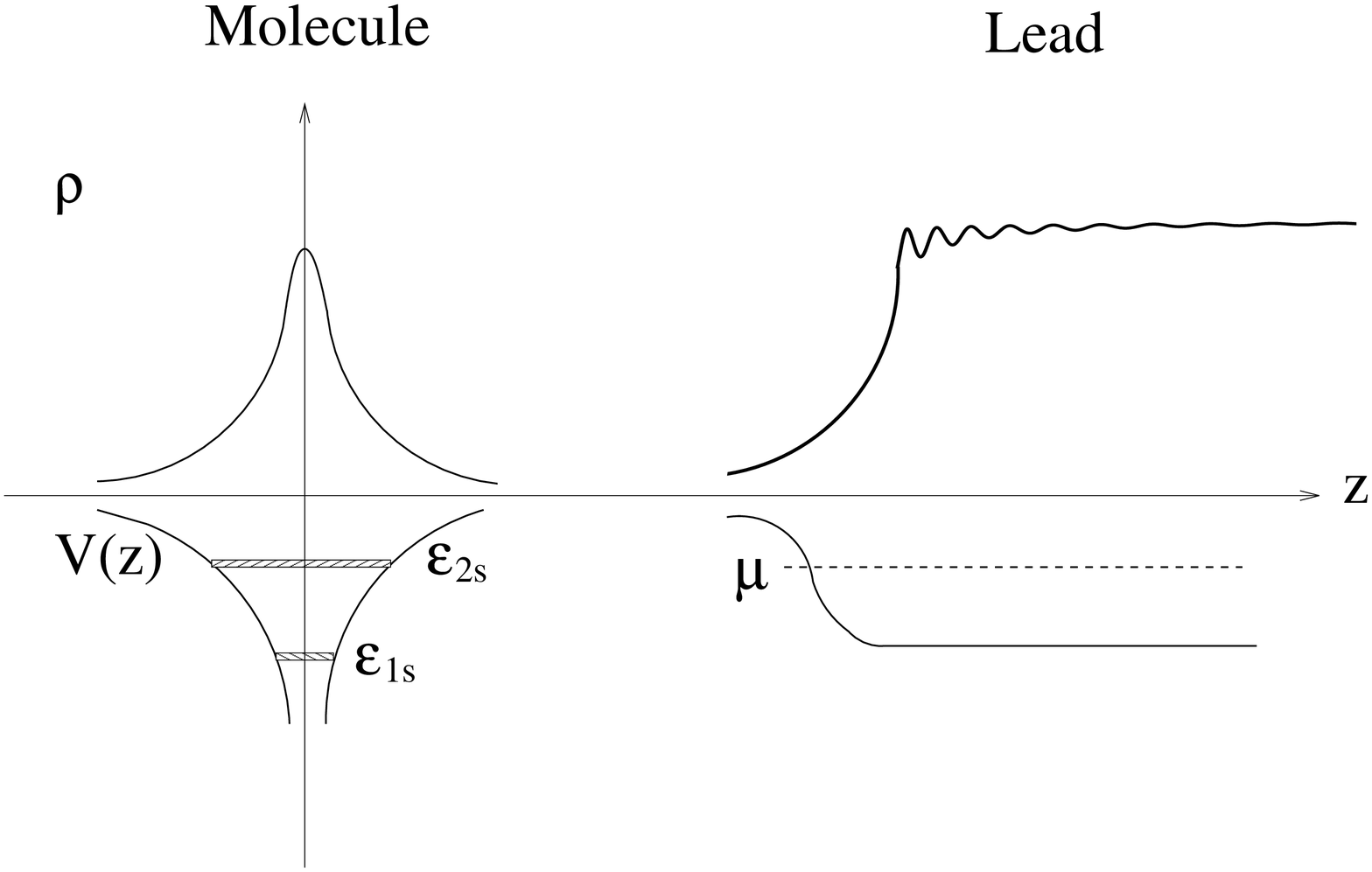}
}}
\end{picture}
\caption{Density and potential of a molecule-lead system where
  resonance occurs when the chemical potential of the lead lines up
  with $\epsilon_{LUMO}$ of the molecule.}
\label{f:cartoon}
\end{figure}

\begin{figure}[tbh]
\begin{center}
\leavevmode
\includegraphics[width=9cm]{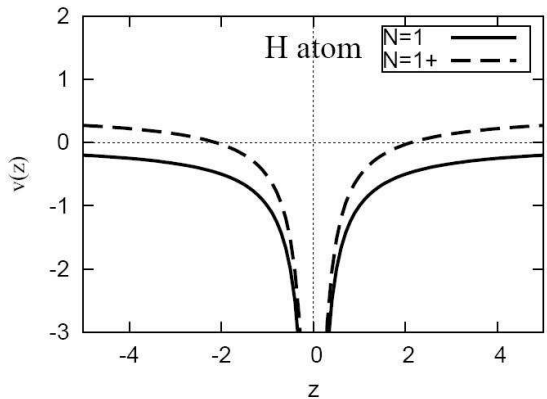} 
\end{center}
\caption{Exact KS potentials for H and H with an extra infinitesimal
  electron, illustrating the derivative discontinuity.}
\label{f:discont}
\end{figure}

What happens then, as electrons are added to the system?  This question
was answered more than 20 years ago.  Consider a Hydrogen atom, far
(say 10\AA) from a featureless metal surface (e.g., jellium), as shown
in Fig.~\ref{f:cartoon},
and ask what happens to the KS potential
as a function of the global chemical
potential of the system.  Since we chose the H atom far from the surface,
its energy levels pick up a tiny width $\gamma$, as they are broadened
into resonances.  If $\mu$ matches the LUMO energy, as shown in the cartoon,
electrons would occupy that level.  But as soon as there's even an infinitesimal
occupation, the level must move, in such a way as to keep the exact
Koopman's theorem satisfied.  Since the density changes at most infinitesimally,
the only allowed change (in the region of the atom) is a constant jump
in the KS potential, by exactly the amount needed to restore Koopman's
theorem for the new HOMO.  This is shown in Fig.~\ref{f:discont}.

\subsection{DFT approximations}
\label{s:gsapprox}

The success of most DFT calculations is based on good approximations to
the XC energy itself, as this determines so many properties of the system.
The most popular approximations, such as LDA\cite{KS65},
GGA\cite{PBE96}, and hybrids of exact exchange and
GGA, such as B3LYP\cite{Bb93}
and PBE0\cite{PEB96}, have well-known successes and failures.
But they have the following failures in common, because they
are {\em density} functionals:

\begin{itemize}

\item
Self-interaction error: none are exact for a one-electron system,
in which 
\ben
E\x=-U,~~~E\c=0~~~~~~~({\rm 1\; electron})
\een

\item
They all have poorly behaving potentials far from the nuclei.
The true KS potential decays as $-1/r$ far from a neutral atom, and this
contribution is from the exchange potential.
\begin{figure}[htb]
\unitlength1cm
\begin{picture}(12,6.5)
\put(-6.5,-4){\makebox(12,6.5){
\includegraphics{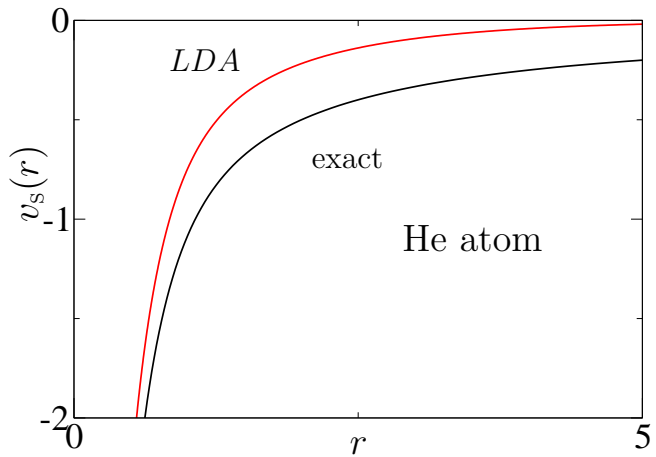}
  }}
\setbox6=\hbox{\Large $v\s(r)$}
\put(0.03,3.5) {\makebox(0,0){\rotl 6}}
\put(4.3,-.1){\Large $r$}
\put(3.8,3.7){\large exact}
\put(1.9,5.0){\large $LDA$}
\put(5,2.6){\Large He atom}
\end{picture}
\caption{Exact and LDA KS potentials for the He atom.}
\label{f:heexact}
\end{figure}
In Fig.~\ref{f:heexact}, we have
also plotted the LDA potential for the He atom.  It decays far too
rapidly, and so its orbitals are far too shallow.  The HOMO is at
-0.5704~H, while the LUMO is not bound at all.

\item
None contain the derivative discontinuity, so their potentials do not
jump when the particle number passes through an integer.

\end{itemize}
(Actually, hybrids have about $1/4$ exact exchange, but this is not enough to cure these ills.)

\begin{figure}[htb]
\unitlength1cm
\begin{picture}(12,6.5)
\put(-6.5,-4){\makebox(12,6.5){
\includegraphics{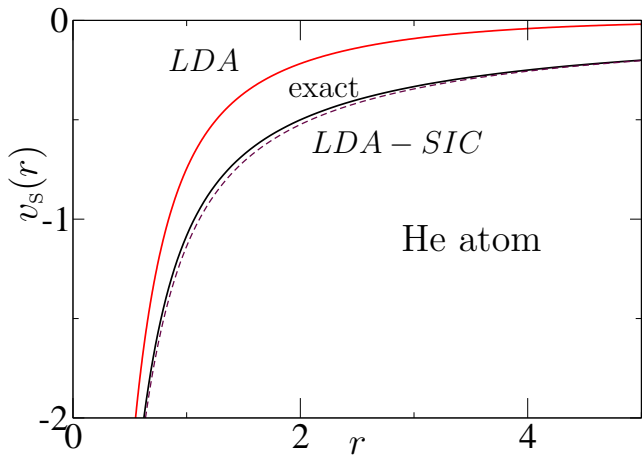}
}}
\setbox6=\hbox{\Large $v\s(r)$}
\put(0.03,3.5) {\makebox(0,0){\rotl 6}}
\put(4.3,-.1){\Large $r$}
\put(3.8,3.9){\large $LDA-SIC$}
\put(3.5,4.65){\large exact}
\put(1.9,5.0){\large $LDA$}
\put(5,2.6){\Large He atom}
\end{picture}
\caption{Perdew Zunger self-interaction corrected KS potential for He
  (dashed line). Exact and LDA KS potentials for the He atom for comparison.}
\label{f:Hesic}
\end{figure}

On the other hand, {\em orbital}-dependent functionals cure all these
ills (at least, approximately).  The original and simplest method
is the self-interaction
correction of Perdew and Zunger\cite{PZ81}, and is often used with LDA for strongly correlated
systems.  The corrected exchange-correlation term is then given by:
\ben
\label{e:LDA-SIC}
E\xc^{LDA-SIC}[n]=E^{LDA}\xc[n]-\sum_i(E_H[n_i]+E\xc^{LDA}[n_i]),
\een
where $n_i(r) =|\phi_i(\br)|^2$.  This functional is exact for
one electron, decays correctly at large $r$, and its potential jumps
discontinuously at integer particle number.
 The dashed line in  Fig.~\ref{f:Hesic} shows the huge
improvement in the potential compared to LDA for the He case.
More satisfactorily, there are now many implementations
of exact exchange within DFT, in which the orbital-dependent
exchange is treated as an implicit density functional\cite{GKKG98}.  Such
optimized effective potential (OEP) calculations are often prohibitively
expensive for large molecules, but
are exact for one electron, have a potential that decays as $-1/r$,
and $v\s(\br)$ jumps discontinuously when a fraction of an electron is added.

Without some form of self-interaction correction, continuous density
functionals allow electrons to self-repel, yielding orbitals that
are too diffuse, in potentials that are too shallow.

\subsection{Effect on transport}
\label{s:efftrans}

The missing orbital effects in approximate density functionals
can have drastic consequences for
calculations of conductances. For example, the missing derivative 
discontinuity in local approximations of DFT affects
the magnitude of the conductance and misplaces the resonance peaks.  This effect
is strongest when the molecule is coupled weakly to the leads. For the exact Kohn-Sham potential as
discussed above, the potential is discontinuous, suddenly shifting
by a constant while the energy remains continuous, as the next unoccupied level
begins to be infinitesimally  occupied (see Fig.~\ref{f:discont}).  The origin of this discontinuity is due
to the fact that $\epsilon_{LUMO}$, the Kohn-Sham
LUMO (lowest unoccupied molecular orbital) for the $N$-electron system is not the
same as the Kohn-Sham HOMO (highest occupied molecular orbital) for the
$N+1$-electron system, as seen in sec.~\ref{s:gsDFT}. In local approximations such as LDA and
GGA, the fractionally occupied level shifts and moves continuously to the
position of the HOMO of the $N+1$ system as the next unoccupied level
begins to be fractionally populated.  The effects for transport calculation are broad
resonance peaks in transmission which lead to finite conductance even
at energies off resonance, yielding an overall increase in conductance that is
unphysical.

\begin{figure}[tbh]
\begin{center}
\leavevmode
\includegraphics[width=8.5cm]{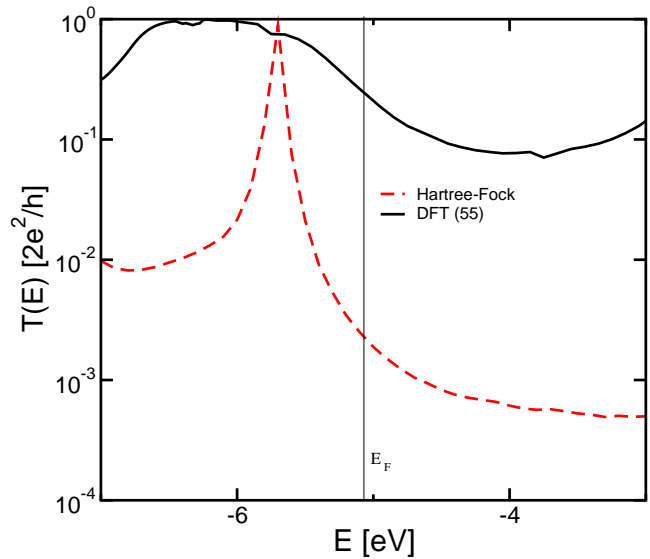}
\end{center}
\caption{Transmission coefficient over energy for benzene-1,4-di-thiol
  using DFT in the {\em standard approach} within a semi-local
  approximation (GGA) (solid line) compared with HF results (dashed
  line). Fermi energy is $\approx -5.1$~eV\cite{EWK04}.}
\label{f:max2}
\label{f:DFT_HF}
\end{figure}

This was first illustrated in ref.~\cite{EWK04}, where Evers et al.\
performed a test calculation on a realistic system, finding
transmission coefficients for benzene-1,4-di-thiol covalently coupled to two gold
clusters using DFT with a GGA,
and comparing to Hartree Fock (HF) methods.
The calculations were performed only at zero bias and the results
for the transmission (found from the corresponding Green's functions) 
are shown in Fig.~\ref{f:max2}.  
The transmission coefficient at the Fermi energy is
reduced by a factor of 100 in the HF calculation.
This is likely due to the GGA orbitals being too diffuse, due to the
self-interaction error.

A more serious problem is due to the incorrect positions of unoccupied levels.
To probe an unoccupied resonance at zero bias, we can apply a gate voltage
$V_g$ to a double barrier resonant-tunneling device (DBRTS)
perpendicular to the leads, shifting the LUMO down to the Fermi level $\epsilon_F$ ($=\mu$ at $T=0$).  In Fig.~\ref{f:barrier}, this simply reduces all molecular levels by $V_g$. 
As a level passes through $\epsilon_F$ as a function of the gate voltage, there will
be a peak in the conductance.
When the resonance starts to overlap with
$\epsilon_F$, the exact KS ground-state potential in the region of the molecule will differ
significantly from the off-resonant situation, as it depends on the
occupation, i.e. it will jump discontinuously by the derivative
discontinuity.
This thereby greatly changes the transmission
characteristics.  The transmission peaks are {\em not} at the
position of  the unoccupied resonances of the ungated situation.

For any sharp resonance, the transmission coefficient is given by
\ben
T(\epsilon)=\frac{(\gamma/2)^2}{(\epsilon-\epsilon\res)^2+(\gamma/2)^2}
\label{T}
\een
where $\epsilon\res$ and $\gamma$ are the position and width
of the resonance.
In any self-consistent KS treatment (including using the {\em exact}
ground-state functional), $\epsilon\R$ and $\gamma$ depend on the
ground-state density, and therefore on the
partial occupation,
$0 \leq f \leq 1 $,
of the resonant level.
\begin{figure}[tbh]
\unitlength1cm
\begin{picture}(8,6.5)
\put(-6.5,10){\makebox(12,6.5){
\includegraphics{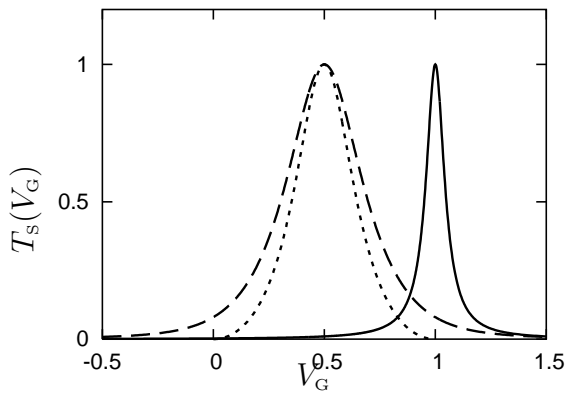}
}}
\setbox6=\hbox{\large $T\s(V_{\sss G})$}
\put(0.4,3.85) {\makebox(0,0){\rotl 6}}
\put(4,1.5) {\large $V_{\sss G}$}
\end{picture}
\vskip -1.5cm
\caption{Double barrier resonant tunneling system with a gate
  electrode. Zero-bias transmission over gate voltage: dashed line is self-consistent
approximate functional result, dotted line is approximate result for
coupling $\gamma\to 0$, and solid line is exact result.  Here $\epsL(N)=0,
\epsH(N+1)=1$ and
$\gamma_0=0.1$\cite{KBE06}.}
\label{f:max}
\end{figure}

We will now use a simple model to show how the use of smooth, approximate density functionals
  produces completely erroneous transmissions (and hence conductances)
  as a function of $V_g$\cite{KBE06}.  Defining the KS spectral function $A\s(\epsilon)=\Im(tr{\bG\s(\epsilon)})$  we can write expressions for the
  spectral density of states, $n(\epsilon)={A\s}/\pi$, as well
  as for the transmission $T\s=\gamma A\s/2$. This yields a simple
  linear relationship between $n(\epsilon)$  and the transmission  of
  such a level, $n(\epsilon)=
  2\, T\s(\epsilon)/(\gamma\pi)$. The self-consistent occupation $f$ of
  the level is found from
  integrating over $n(\epsilon)$ as
\ben
f = \int_{-\infty}^{\epsilon_F} d\epsilon\ n(\epsilon)
= \half+\frac{1}{\pi}\tan^{-1} 
\left\{2 \frac{\epsilon_F-\epsilon\res(f)}{\gamma(f)}\right\}
\label{fdef}
\een
The transmission can be obtained by inverting Eq.~(\ref{fdef}):
\ben
T^{-1}(\epsilon_F) =
1 + \tan^2 \left\{\pi(f(\epsilon_F)-1/2)\right\}.
\label{Tf}
\een
For simplicity, neglect any dependence of $\gamma$ on
occupation $f$, i.e., $\gamma(f)=\gamma_0$, as the actual
dependence is expected to be weak and to have only little effect on the
transmission peaks.
The transmission can alternatively be expressed in terms of the gate voltage.
Setting $\epsilon\F=0$ for $V_g=0$ and assuming a shift of the energy
levels by $-V_g$ due to the applied gate voltage (gate efficiency=1),
we can replace  $\epsilon\F$ by $V_g$ in
Eq.~(\ref{fdef}) and Eq.~(\ref{Tf}), thus describing the transmission
in terms of an applied gate voltage instead:
\ben
T^{-1}(V_g) =
1 + \tan^2 \left\{\pi(f(V_g)-1/2)\right\}.
\een

Any calculation that has a derivative discontinuity would give the solid
line in Fig.~\ref{f:max}.
In this example $\Delta \epsilon=\epsH(N+1)-\epsL(N)$ is several eV. 
The narrow resonance (width $\gamma=0.1$) is positioned at
the energy of $\epsH (N+1)$.  On the other hand, 
the dashed line is the result for a
smooth, (semi-)local density functional. As the $N+1$ level gets fractionally
filled, the resonance moves continuously from  $\epsL (N)$
to $\epsH (N+1)$, resulting in a smearing out of the resonance
width.
It can be seen that the position of the resonance is displaced in this
case. It is now centered inbetween the
LUMO of the $N$ electron system and the HOMO of the $N+1$ electron
system, assuming $\epsilon\res=\epsL +f\Delta \epsilon$ (i.e. a linear dependence of the potential
on the occupation for the smooth functional).  The resonance peak
should be located at the {\em true}  HOMO of the $N+1$ electron system
(the solid line).

Even in the extreme limit of no width of the level ($\gamma \to
0$), the resonance in LDA remains broad. For weakly coupled
leads where, at any occupation,  $\gamma \ll \Delta\epsilon$, the
Fermi level is pinned to the resonance ($\epsilon\res (f) \to
\epsilon_F$) for $f \neq 0$ or $1$. This yields $\epsilon_F=\epsL+ f \Delta
\epsilon $ and  using Eq.\ (\ref{Tf}) we obtain the dotted line in
Fig.~\ref{f:max}.
Thus, in a {\em standard} calculation using (semi)local functionals, Eq.\ (\ref{Tf}) always produces
a broad peak whose width is comparable to $\Delta \epsilon$, thereby
overestimating the total conductance of the device --- even when 
$\gamma\rightarrow 0$.
For the case of a linear relation as discussed here, this artificial width is just $\Delta\epsilon/2$.
In addition, the resonance position is incorrect, being $\Delta\epsilon/2$ too 
low.

It is possible to recover the derivative discontinuity and so avoid these artifacts with methods
briefly described in section \ref{s:gsapprox}. The
self-interaction is effectively removed either approximately from an
LDA description via a self-interaction correction (LDA-SIC) or through
the rather expensive, but more rigorous, OEP methods. In this method, the self interaction can be removed 
explicitly, but it is computationally more feasible to parameterize the
self-interaction in terms of its atomic components.  What results is an
orbital-dependent functional that incorporates non-local effects,
yielding the derivative discontinuity.

\begin{figure}[htb]
\unitlength1cm
\begin{picture}(10.8,4.1)
\put(-5.7,-4.8){\makebox(12,9){
\includegraphics{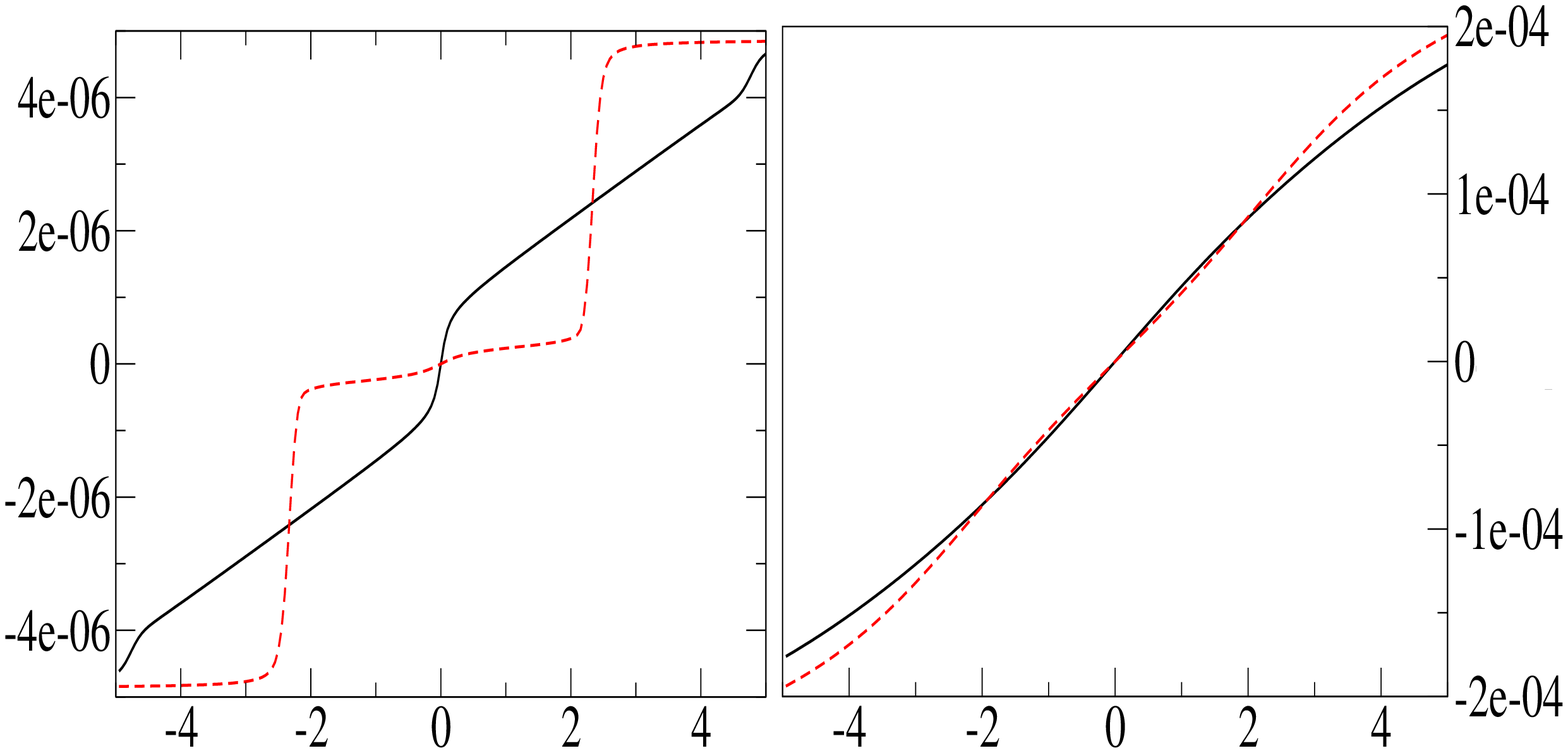}
}}
\put(0.2,5.9) {\makebox(0,0){\rotl 6}}
\setbox6=\hbox{\large $I(A)$}
\put(8.9,2) {\makebox(0,0){\rotl 6}}
\setbox6=\hbox{\large $I(A)$}
\put(0.3,2.) {\makebox(0,0){\rotl 6}}
\put(1.9,-.8){\large $V(volts)$}
\put(5.8,-.8){\large $V(volts)$}
\put(3.5,2.0){\large $SIC$}
\put(1.9,2.4){\large $LDA$}
\put(6.6,3.2){\large $SIC$}
\put(7.2,2.4){\large $LDA$}
\end{picture}
\vspace*{0.3cm}
\caption{Current of a single energy level coupled to two metallic leads
  as a function of bias.  Left figure corresponds to the case of weak coupling
  and right figure corresponds to the case of strong coupling. Solid
  lines are results for LDA and the dotted lines are results using
  self-interaction corrected LDA (LDA-SIC). From Ref.~\cite{TFSB05}.} 
\label{f:Sanvito1}
\end{figure}

Comparison of I-V curves between LDA and LDA-SIC was performed by
 Toher {\it et.~al}\cite{TFSB05,TS07} using a
tight-binding calculation within the {\em standard approach}.  The differences in the I-V
characteristics were much less apparent for the case of strong coupling,
confirming that the missing derivative discontinuity is most problematic in the
limit of weak coupling.  A cartoon of this effect can be seen in
 Fig.~\ref{f:Sanvito1}
where the current versus applied bias voltage of a
device effective single-particle 
energy level is plotted for the LDA case (solid line), and the case
for an
artificial step-like energy (dotted line) which emulates the derivative discontinuity for the
exact Kohn-Sham potential.  The plot on the left hand side reflects the
situation of weak coupling to the leads ($\gamma=0.2$eV), whereas the plot on the right hand side
reflects the situation of strong coupling to the leads( $\gamma=1.2$eV). 

\section{Weak bias}
\label{s:weak}
In this section, we discuss --- in the weak bias limit --- the errors made
due to the {\em standard approach} scheme.  In this limit, we can use Kubo response
theory to deduce the exact answer, and compare with the {\em standard approach}.  We also give an
estimate of the corrections in terms of the Vignale-Kohn current
density functional.
Local functionals not only miss the derivative
discontinuity but also the XC contributions to the electric field response,
which leads to an additional overestimation of the conductance.    
In the limit of low bias it is possible to describe transport with the Kubo response
formulation.  Seeing how the adiabatic local density approximation
(ALDA) fails in this formulation provides clues as to the
source of the problem and how to correct it.  A more detailed
discussion of calculations in this limit is given in
section~(\ref{s:diffchem}).

\subsection{Time-dependent DFT}
\label{s:tddft}
The usefulness of ground-state DFT has been augmented by
the development and implementation of TDDFT\cite{BWG05}.  The
Runge-Gross theorem\cite{RG84} shows that, under appropriate conditions, the time-dependent
potential $v\ext(\br,t)$ is a functional of the time-dependent density, $\n(\br,t)$.
This allows one to construct time-dependent Kohn-Sham equations,
and use linear response theory to find TDDFT corrections
to the KS transitions, making them the true excitations
of the system.

How the 
linear response theory works can be seen in the density change to
an applied perturbation varying as $\exp(i\omega t)$, which can be
expressed {\em exactly} in different ways:
\bea
\delta\n(\br\omega)&=& \int d^3r\, \chi(\br,\br',\omega)\, 
\delta v\ext(\br,\omega)~~~~~({\rm MB})\nonumber\\
&=& \int d^3r\, \chi_{prop}(\br,\br',\omega)\,
\delta v\tot(\br,\omega)~~~~~({\rm EM})\nonumber\\
&=& \int d^3r\, \chi\s(\br,\br',\omega)\,
\delta v\s(\br,\omega)~~~~~({\rm DFT})\nonumber
\eea
where $v\tot (\br\omega)$ is the sum of the external and induced
(a.k.a.\ Hartree) potentials, while 
\ben
\delta v\s(\br,\omega)=
\delta v\ext(\br,\omega) + \delta v\H(\br,\omega) 
+\delta v\xc(\br,\omega)
\een
is the Kohn-Sham potential perturbation including the XC contribution.
Different susceptibilities are used in different contexts:  $\chi$
is the full many-body (MB) response function, giving the density change in
response to the {\em external} perturbation; $\chi_{prop}$ is the
proper or irreducible susceptibility, giving the response to the perturbing potential
of the total electric field, both external and induced (Hartree), used
in electromagnetism (EM)\cite{M90}.
Finally, $\chi\s$ is the {\em Kohn-Sham} response function, constructed
from KS energies and orbitals of the ground-state KS potential:
\ben
\chi\s(\br\br'\omega) = 2 \sum_{q}
\frac{\Phi_q(\br)\ \Phi_q^*(\br')}
{\omega-\omega_q+i 0_+} + c.c.(\omega\to-\omega),
\label{e:chis}
\een
and
\ben
\langle q | f | q' \rangle = \int d^3 r\, \int d^3 r' \,
\Phi_q (\br) f(\br,\br') \Phi_{q'}(\br')
\een
where $q$ is a double index, representing a transition from
occupied KS orbital $i$ to unoccupied KS orbital $a$,
$\omega_q=\epsilon_a-\epsilon_i$,
and $\Phi_q(\br) = \phi_i^*(\br)\phi_a(\br)$, where $\phi_i(\br)$ is
a KS orbital.  Thus $\chi\s$ is
completely given by the ground-state KS calculation.
For example, it is the response of the two non-interacting KS electrons sitting in the KS potential
of Fig. 2, and $\omega_q$ are the differences of the orbital energies listed in Table~\ref{t:heKS}.
By definition, $\delta v\xc(\br,\omega)$ causes these non-interacting
electrons to have the same density response as the real electrons.
Expanding this around the original ground-state density, and requiring
the same density response, we find a Dyson-like equation relating the
true and KS susceptibilities\cite{GK85}.  This is just the RPA equation well-known
in other areas, but with the Hartree interaction modified to include XC
effects.  One can further translate this problem into an eigenvalue problem\cite{C96},
whose approximate solution for transition frequencies is \cite{AGB03}:
\ben
\omega \approx \omega_q + 2 \langle q | \frac{1}{|\br-\br'|}+f\xc |q \rangle
\een
where 
\ben
f\xc[\n_0](\br\br',t-t') = \delta v\xc(\br t)/\delta \n(\br' t')|_{\n_0}.
\label{e:fxc}
\een
is called the XC kernel.  Thus, the effect of TDDFT is to produce corrections
to the KS transitions to turn them into the true optical transitions of the
system. 
If we had the {\it{exact}} $f\xc(\br,\bf{r'},\omega)$ for the He atom density, calculated perhaps from a traditional wavefunction calculation, and inserted it in the full TDDFT response equations, we would get {\it{exactly}} the results of 
Table~\ref{tab:hetrans}.  
\begin{table}
\caption{\label{tab:hetrans} Singlet transition energies for He, comparison of
  the true transitions with the Kohn-Sham transitions for the exact KS
  potential\cite{FB06}.}
\begin{ruledtabular}
\begin{tabular}{cdd}
  \multicolumn{1}{c}{transition} & \multicolumn{1}{c}{KS value [eV]\footnotemark[1]}&
  \multicolumn{1}{c}{true transition [eV]\footnotemark[2]}\\\hline  
 $1s\rightarrow 2p$ & 21.146 & 21.221\\
  $1s\rightarrow 2s$ & 20.298 & 20.613 \\
    $1s\rightarrow 3s$ & 22.834  & 22.923
      
\footnotetext[2]{The differences between the KS eigenvalues obtained 
using the exact potential \cite{FB06}.}
 \footnotetext[2]{Accurate non-relativistic calculations from 
Ref.\cite{Dc96}.}
\end{tabular}
\end{ruledtabular}
\end{table}

\subsection{TDDFT approximations}
\label{s:tddftapprox}

Most often, ground-state approximate XC-functionals
are used, even for the TD potential, which is called the
adiabatic approximation.   This often produces excellent excited-state
properties, and transition frequencies typically within about 0.2 eV
of the true numbers for molecules\cite{FR05}.
A simple example is the $\pi\to\pi^*$ transition
in benzene in which, in a LDA calculation, the KS transition is
at about 5 eV, but the TDDFT (ALDA) correction correctly shifts it to about 7 eV\cite{VOC02}.

TDDFT has been implemented in many standard quantum chemistry
codes, and is run routinely to extract electronic excitations of
molecules\cite{FR05}.   However, as the number of calculations has grown very
rapidly, the limitations of the scheme with an adiabatic XC approximation
are being felt.  Even early on\cite{PGG96}, it was realized that Rydberg excitations
would be missing within ALDA or AGGA, because of the poor quality potentials
of the underlying ground-state approximation.  The LDA potential
of Fig.~\ref{f:heexact} is shallow and short-ranged, and so has no Rydberg series.  Exact exchange or SIC functionals
take care of this \cite{FB06}; Fig.~\ref{f:Hesic} shows how accurate
the LDA-SIC potential is in comparison.  Also, double-excitations can be shown to be missing within
any adiabatic approximation\cite{C96}, although frequency-dependent kernels can be constructed
that restore them\cite{MZCB04}.  
Charge transfer-type excitations also
fail\cite{MT05}.

For solids, development has been slower, as the
local and semi-local nature of approximate functionals means that
their effect becomes negligible in the thermodynamic limit.
This can be seen from the fact that the XC kernel in ALDA is independent
of $q$, the wave vector in the Fourier transform of $\br-\br'$, as $q\to 0$, but
the Hartree term grows as $1/q^2$.   One cure for this problem
is to use TD current DFT (TDCDFT)\cite{BKBL01}, whose validity is established in the first part
of the RG theorem\cite{RG84}.  Related to this fact is the notion that no local
approximation exists in terms of the density, but a gradient expansion in 
the current was
constructed by Vignale and Kohn\cite{VK96} for linear response.  More recently,
by comparison with solutions of the Bethe-Salpeter equation, accurate
many-body approximations to the kernel have been constructed, yielding
excellent results for excitonic peaks, etc\cite{MUNR06}.

The VK approximation is the only current-dependent approximation that is well-established, and is often applied to problems where non-locality and
memory (i.e.\ beyond adiabatic) effects are important.  Most importantly, it often yields finite
corrections where ALDA gives nothing, such as to the (0,0) component
needed for the optical response of
solids\cite{BKBL01,MSB03}, or the over-polarizability of long-chain 
conjugated polymers\cite{FBLB02,FBc04}.
But, given that it is a simple gradient expansion, its
quantitative accuracy in any given situation is open to question\cite{FBb04,FBc04,UB04}.

\subsection{Constitutive relations}
\label{s:directresponse}

We can relate the conductivity $\sigma$ with the susceptibility $\chi$
described in section~(\ref{s:tddft}) using  current continuity
\ben
{dn\over{dt}}=-\nabla\cdot\bj.
\een
Since $\delta n(t) = \delta n e^{i\omega t} $, we have:
\ben
\delta n = -{1\over{i\omega}}{\nabla}\cdot\delta\bj.
\een
Then, using the relations between the vector potential and the field:
\ben
\delta \bE  = {\partial{\delta\bf{A}}\over{\partial{t}}}  =  i\omega
\delta \bf{A},
\label{e:relpotfield}
\een
and the definitions of the current response:
\bea
\label{e:currentresponsedef}
\delta j_\alpha(\br\omega)&=&\int{d^3\br'}\sum_{\beta}\chi_{\alpha
  \beta}(\br\br'\omega)\delta A_\beta(\br'\omega)\cr
&=&\int{d^3\br'}\sum_\beta\sigma_{\alpha \beta}(\br\br'\omega)\delta E_\beta(\br'\omega),\cr
&&\eea
where $\chi_{\alpha \beta}$ is the current-current response function, a tensor, and 
$\sigma_{\alpha \beta}$ is the conductivity tensor\cite{UV02}.  
We get an expression relating the  response functions for conductivity and susceptibility:
\bea
\sigma_{\alpha \beta}={1\over{i\omega}}\chi_{\alpha \beta}
\eea
Also, going back to the definition of the density response
and using the relations given above in equations
(\ref{e:relpotfield}-\ref{e:currentresponsedef}), we obtain:
\ben
\delta n(\br\omega)={1\over{\omega^2}}\int{d^3r}
\sum_\beta\partial_{\alpha}\sigma_{\alpha \beta}(\br\br'\omega)(-\partial'_\beta v(\br\br'\omega))
\een
which leads to the relation between the conductivity and susceptibility:
\ben
\sum_{\alpha \beta}\partial_\alpha\partial'_\beta \hat{\sigma}_{\alpha
  \beta}(\br\br'\omega)=-i\omega\chi(\br\br'\omega)\label{e:condsusc}
\een

As described in section \ref{s:tddft}, in TDDFT, there are three
equivalent, exact formulas
-- usually studied in reference to the polarizabilities of atoms in
strong fields -- applied to different many
body descriptions and requiring different inputs, to describe the density
response to an applied electric field.

There is an analogous response function for the current response to an external
electric field.  As with the density response formula, the response,
given by the conductivity $\sigma$ in this
case, is different in each of the  exact expressions.  The many-body,
non-local conductivity $\sigma(\br \br'\omega)$ describes the response to the external
electric field $\delta \bE\ext$.  The proper conductivity $\sigma_{prop}$ is in
response to the total field $\bE\tot=\bE\ext+\bE\H$, and the single particle Kohn-Sham
conductivity $\sigma_{S}$ yields the response to the full expression for the electric
field, $\bE\ext+\bE\H+\bE\xc$, including the unphysical XC contributions $\bE\xc$.    
\bea
\delta \bj(\br\omega)&=& \int dr\, \sigma(\br,\br',\omega)\, 
\delta \bE\ext(\br,\omega)\, ({\rm MB})\\   
 &=& \int dr\, \sigma_{prop}(\br,\br',\omega)\,
\delta \bE\tot(\br,\omega)\;  ({\rm EM})\nonumber\\
&=& \int dr\, \sigma\s(\br,\br',\omega)\,
\delta \bE\s(\br,\omega).\;  ({\rm DFT})\nonumber
\label{e:currentresponse}
\eea

\subsection{DC transport from Kubo response}
\label{s:dcTrans}
In the limiting case of weak bias,
the response can be expanded to first order in the electric
field.  In this, we follow the logic of Kamenev
and Kohn \cite{KK01}.  To derive the DC transport response, a
frequency-dependent electric
field is applied, and the limit $\omega\to 0$ is taken while always ensuring 
that $v_F/\omega << L$, where $L$ is the circumference of
the ring.   As shown by Kamenev and Kohn, this reproduces the
Landauer formula for weak bias for Hartree-interacting
electrons.   Our work can be regarded as a simple
extension of this analysis to DFT.  Note that extreme care must be
taken in the limiting procedure to extract the relevant results\cite{BJG06,BJG07,JBG07}.

It can be shown that, as $\omega \to 0$, the conductivity can be rewritten as the
transmission coefficient familiar from the Landauer formulation.
This limit has to be performed carefully to obtain the correct current.
It has to be assured that the excursion length of electrons in the device, given
by  $l\F=\frac{v\F}{\omega}$, is smaller than the region of the
density response, $l_{\rho}$, which in turn has to be smaller than the device
dimensions $L$ of the extended molecule in the DFT calculation (see
Fig.~\ref{f:benzene}).

Using the (DFT) response  equation for the Kohn-Sham susceptibility
and the full expression for the field we can obtain the correct current.
We first rewrite the Kohn-Sham non-local conductance as
 \ben
\hat\sigma\s(\br\br'\omega) = \left(
\n_0(\br)\;\delta^{(3)}(\br-\br')\; \openone
+\hat R (\br\br'\omega)
\right)/(i\omega)
\een
where
\bea
\hat R(\br\br'\omega)&=&\half\; \sum_{q}
\frac{\bP_q(\br)\bP_q^*(\br')}
{\omega+\omega_a+i 0_+},
\label{chijs}
\eea
defining ${\bP}_q(\br) = \phi_i^*(\br)\nabla \phi_j(\br)-\phi_j(\br)\nabla \phi_i^*(\br)$.
Here
 $\phi_i(\br)$ and $\epsilon_i$ are the KS orbitals
and eigenvalues, $\omega_a=\epsilon_i-\epsilon_j$, and $q=(i,j)$.
This result can be written more compactly in terms of the
retarded KS Green's function $ G^r\s(\br\br'\epsilon) $ and the corresponding KS spectral density
\ben
\label{e:specdens}
A\s(\br\br'\epsilon) = -\Im G^r\s(\br\br'\epsilon)/\pi,
\een
just as the regular $\chi\s$ can be.
Thus we find, exactly,  
\bea
\hat R(\br\br'\omega)&=&\frac{1}{2}\int d\epsilon f(\epsilon)
\{ G^r\s(\br\br',\epsilon+\omega) \\
& & +  (G^r\s)^*(\br\br',\epsilon-\omega)\}
\tensor\nabla \tensor\nabla' A\s(\br\br'\epsilon)\nonumber
\eea

For small $\omega$, only the imaginary component of the KS Green's function
contributes to $\hat R$. Expansion in powers of $\omega$ yields a term linear in $\omega$, and an integration by parts yields the DC
conductance entirely in terms of the spectral density at the Fermi energy\cite{KK01}:
\ben
\hat\sigma\s(\omega\to 0) = -\pi
 A\s(\epsilon_F,\br, \br') \tensor\nabla \otimes \tensor\nabla'
A\s(\epsilon_F,\br, \br').
\label{sigmafromA}
\een
This result is true for the conductance of non-interacting electrons
in any single-particle potential.
Next, we specialize to a 1d system, to avoid complications. Then, Eq.~(\ref{e:condsusc}) tells us that, as $\omega \to 0$, $\sigma\s$ becomes independent of position. Thus, $\sigma\s$ from Eq.~(\ref{sigmafromA}) may be evaluated at any choice of $z$ and $z'$. An easy choice is $z < 0$ and $z' > 0$, and one finds \cite{KK01}
\ben
\sigma\s(\omega \to 0) = \frac{T\s(\eps\F)}{\pi}.
\een
Since $\sigma\s(\omega \to 0)$ is just a constant, it can be taken outside the integral of Eq.~(\ref{e:currentresponse}), yielding

\bea
\label{e:I1}
\delta I&=&{T\s\over{\pi}}(\delta V\tot + \delta V\xc)\\
\delta I(\omega\to 0)&=&{{T\s(\epsilon\F)}\over{\pi}}\int d^3r'(\delta
E_{ext}(\omega)\\
& +&\delta E\H(r'\omega)+\delta E\xc(r'\omega))
\eea
 where  $\delta V\tot=\int dz'\, \delta E\ext(\omega) + \delta E\H (z'
 \omega)$ is the net drop in total electrostatic potential across the
 device, and
 \ben
\delta V\xc=\int_{-\infty}^{\infty} dz\, \delta E_{\sss z,XC}
 (z,\omega\rightarrow 0)
 \label{e:vxcterm}
\een
 is the corresponding drop (if any) in the XC-potential.

 Thus, Eq. (\ref{e:I2}) would be exact if the
exact $V\xc$ was properly included.
But the implementation commonly used in the Landauer
formulation of molecular electronics corresponds, as we will see below
in \ref{s:xccorr}, to only the Hartree response:
\bea
\label{e:I2}
\delta I&=&{1\over{\pi}}\int^{\mu+\delta V}_{\mu}d\epsilon\,
T\s (\epsilon,V)(f_{\sss
  L}(\epsilon)-f_{\sss R}(\epsilon))\cr
 &=&{T\s (\epsilon\F)\over{\pi}}\, \delta V\tot\;\;\;\;({\rm LANDAUER})
\eea
Equations~(\ref{e:I1}) and~(\ref{e:I2}) are identical,
except that the standard approach does not include the extra exchange-correlation
 term, $\delta V\xc$.  This derivation has been recently generalized to
include correct averaging over the lateral directions\cite{PC07}.

\subsection{XC correction to current}
\label{s:xccorr}
The present implementation of the Landauer 
formulation using local functionals includes the Hartree piece of
the potential and thus correctly includes the charging effects, but it is
missing the XC piece.  To see this, consider the XC contribution to the voltage
given by Equation (\ref{e:vxcterm}).
Since $\delta E\xc=-\bigtriangledown\cdot\delta v\xc$, this implies
that $\delta V\xc=\delta v\xc(z\rightarrow\infty)-\delta v\xc(z\rightarrow
-\infty)$.  But far from the barrier, $\delta \rho = 0$,
and so any local or semi-local approximation necessitates that
  $\delta v\xc$ equals zero far from the barrier. Thus $\delta V\xc=0$ when working
within these approximations~\cite{KBE06}.  Thus ALDA and all other
local or semi-local approximations miss the non-local interactions of the exact XC functional.

Alternatively, integration of 
the second equation in expressions (\ref{e:currentresponse}) would also give the exact result since all
  three formulas are equivalent and exact:
\ben
\label{e:exactcurrent}
\delta I={T_{prop}\over{\pi}}\delta V\tot
\een
but $T_{prop}$ refers to the full proper transmission coefficient which
cannot be easily calculated for a realistic system, and is not the
transmission through any single-particle potential.

TDDFT within a local or semi-local approximation has been shown to produce erroneous
results when non-local effects
become important, such as in the optical response of solids.
The Vignale-Kohn functional is
a non-local functional in terms of current density that has been successfully
applied to situations where non-locality cannot be ignored, such as 
in long conjugated polymers\cite{FBLB02,FBLB03,FBc04} where local
approximations give overestimates on the static polarizabilities.  This
non-locality also plays a role in the non-equilibrium transport problem as
seen in section\ref{s:dcTrans}. 
In the regime of weak bias, Koentopp {\it et al.} \cite{KBE06} estimate the size of the XC correction to the
current in the Vignale-Kohn approximation.  Since ALDA is a local approximation,
it misses the non-local interactions of the exact XC functional.  Inclusion of
the viscous contribution to $\delta V\xc$
yields a correction to the transmission coefficient that reduces its magnitude:
\ben
\delta V\xc /V\approx-(1-T(\epsilon\F))T(\epsilon\F)/40\pi^{1/2}k\F^{3/2}
\een

A more explicit expression for the XC
correction can be calculated.  
Sai {\it et al.} \cite{SZVD05} calculate the dynamical response contribution to current flow using
the Vignale-Kohn correction in TDCDFT.  This dynamical contribution is a viscous
flow component from the XC field that is missing in ground state DFT
calculations and gives a finite correction to the conductance.  A current
density functional theory is necessary because functionals that are dependent
on the density alone don't contain information about the constant value of the
current.  The Vignale-Kohn construction has an XC field that has both the
ALDA XC potential and a term dependent on the XC stress tensor which in turn is
dependent on viscoelastic coefficients and velocity fields.   

A dynamical resistance $R^{dyn}$ arises from the DC XC field which increases the
total resistance of the system, thus acting against the external field.
\ben
\label{e:dynR}
R^{dyn}={4\over{3e^2A_c}}\int^{\infty}_{-\infty} \eta {(\partial_z n)^2\over{n^4}}dz
\een   
where $a$ and $b$ are points inside the electrodes, $A_c$ is the cross sectional
area, $n$ is the density, and $\eta$ is the viscosity.  A calculation that includes the real part of the stress component of the electric field yields a correction of $10\%$\cite{SZVD05}.

\section{Finite bias}
\label{s:finitebias}

Given the limitations of the {\em standard approach} already discussed, 
it has been realized that a more fundamental derivation of the
conductance formula is needed, especially one that lends itself
to a DFT treatment.  For example, in DFT, one is not allowed to 
turn off the coupling between the molecule and leads, as the
Hohenberg-Kohn theorem does not apply in empty space, nor does
the RG theorem allow for time-dependent interactions between electrons.

Several suggestions have been made as to how to do this, that might
appear quite different.  Here we discuss and compare just two of these:
the Master Equation approach, and the TDDFT-NEGF approach.  
The Master Equation requires coupling to a dissipative bath, such as the
phonons, in order to achieve a steady current, while the TDDFT-NEGF 
approach achieves a steady current via dephasing into the continuum.
Moreover, the Master Equation allows for periodic boundary conditions (PBC's)
whereas TDDFT-NEGF uses a localized system.  Finally, because of this,
the Master Equation with periodic boundary conditions requires TD current DFT, whereas TDDFT-NEGF uses the
density as the basic variable.

\subsection{Master Equation}
\label{s:Master}

\subsubsection{Periodic Boundary Conditions}
\label{s:diffchem}
The Landauer formulation, and indeed most of
the literature on transport, uses the concept
of different chemical potentials on the left- and right-electrodes, and
assumes some steady
current-carrying state between them.

Such a situation is not so easy to realize within the basic theorems of
density functional
theory, time-dependent or otherwise.  Even
TDDFT requires starting from some initial
wavefunction\cite{MB01}, almost always the ground-state
wavefunction of some system.  But the ground
state of any system of electrons has at most
one chemical potential, not two.

Thus useful DFT descriptions begin with a
system in its electronic ground-state, and
a single chemical potential.  So far, only the situation in
which both electrodes are of the same metal have been investigated.
Furthermore, to avoid the difficulties of
having infinite potentials far inside the
electrodes, a gauge transformation that is
standard in solid-state physics is applied and
a solenoidal magnetic field is imposed on a ring of
material.  A vector potential that is linear
in time, $a=E t$ then gives rise to a uniform electric field on the ring,
causing
a current to flow.

\begin{figure}[tbh]
\begin{center}
\leavevmode
\includegraphics[angle=270,width=10cm]{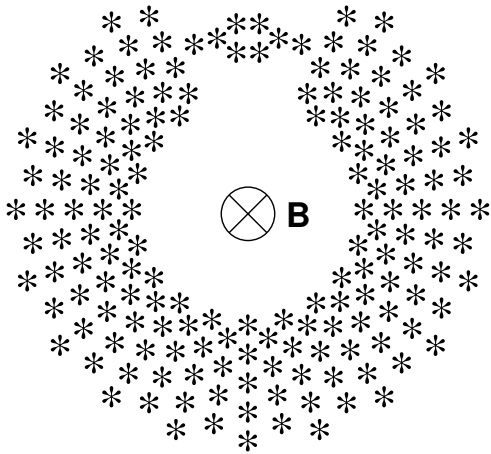}
\end{center}
\caption{Master Equation schematic - periodic boundary conditions, magnetic
  field induces electric field on ring.}
\label{f:fig2}
\end{figure}

The same approach is then applied to finite bias, and
again avoids the need for two different chemical potentials.
However, a new complication arises, as in
the presence of the electric field, as $\omega\to 0$, if
$L$ is kept finite, the electrons will accelerate indefinitely
around the ring, which is not the situation we wish to
model.  Instead, if some coupling to the phonons in the
system is introduced, there will be dissipation, and a
steady state can develop.  It is possible to derive an extension of 
TDCDFT that includes dissipation in a
time-dependent Kohn-Sham Master Equation.  Note that
dissipation is unnecessary in the weak bias limit
of the previous section, as Joule heating is second-order
in the perturbation\cite{KL57}.

\subsubsection{Master Equation theory}
\label{s:mastertheory}
 A master equation approach has been constructed that introduces dissipation
via a quantum mechanical treatment of the Boltzmann equation from statistical
mechanics\cite{BCG05}.  The master equation describes the evolution of the density of a
system coupled to a heat bath and has an analogue with the well-studied problems
of optical interactions of laser fields with matter\cite{CDG92} -- atomic transitions in
the presence of electromagnetic fields.  Among the advantages of this approach is the elimination of the artifical
boundary and contact conditions necessary in the Landauer formulation, the
inclusion of inelastic processes, and applications beyond the steady state
situation.  

With this approach, there are no reservoirs corresponding to the left and
right leads at different chemical potentials and 
the voltage drop across the barrier is an output of the method rather than an
input as it is for the Landauer approach.
Instead of the open boundary conditions employed by the Landauer approach as
illustrated in figure (\ref{f:fig1}),  periodic boundary conditions are imposed
\cite{BCG05} such that the system is a
closed circuit with no exchange of electrons with semi-infinite reservoirs (see
Fig.~\ref{f:fig2}).  This geometry is a neat way
to treat an open system, avoiding partitioning an infinite system as
in the {\it standard approach}.

The approach employs a quantum Liouville equation for the total Hamiltonian
of the system $H_{T}$, which contains the device Hamiltonian $H_{0}$, the phonon bath $R$, and the electron-phonon coupling
potential $V$. $S_{T}$ is the density
matrix for the total system. Its time evolution is given by
\ben
\label{e:first}
 dS_{T}/dt=-i[H_{T},S_{T}].
\een

After tracing out the bath degrees of freedom, what results is the Liouville
equation for the
reduced density matrix $S$ along with a term that encapsulates
the dissipation in the system $C[S]$:
\ben
\label{e:mastereq}
dS/dt=-i[H,S]+C[S].
\een

The dissipation term describes collisions with the heat bath and is given by
\bea
C[S] & = & -\sum_{m,n}\Gamma_{mn}(L_{nm}L_{mn}S+SL_{nm}L_{mn}
\nonumber\\
  & &    -2L_{mn}SL_{nm}),
\eea
where $L=c_{n}^{\dagger}c_m$ and $\Gamma_{mn}$ are the transition
probabilities obtained through Fermi's golden rule. To derive explicit
expression, the coupling potential
\ben
V=\sum_{m,n,\alpha}\gamma_{m,n}^{\alpha}c^{\dagger}_nc_ma^{\dagger}_{\alpha}
+ h.c.
\een
is treated  perturbatively to second order. 
 The creation and destruction operators for the electrons/phonons are $c^{\dagger}/c$
and $a^{\dagger}/a$ respectively. In the coupling matrix elements,  $\gamma^{\alpha}_{m,n}$, the indices
$m$ and $n$ refer to the electrons and $\alpha$ refers to the phonons. 
$\Gamma_{mn}$ is then given by:
\bea
\label{e:dissip}
D(\epsilon_n-\epsilon_m)|\gamma_{mn}|^2(\bar{n}_{\epsilon_n-\epsilon_m}+1),
  \,\epsilon_n > \epsilon_m\cr
D(\epsilon_m-\epsilon_n)|\gamma_{mn}|^2\bar{n}_{\epsilon_m-\epsilon_n}\,,\epsilon_m
  > \epsilon_n. 
\eea
The electric field is imposed on the system through the addition of a time
dependent vector
potential $a(t)$ in the Hamiltonian.  Gauge transformations are then performed
periodically to set the vector potential to zero, otherwise the Hamiltonian
would grow indefinitely leading to numerical instability in the implementation.  
One of the problems with previous attempts to use a Master Equation
formulation within this setup is the apparent current continuity violation. 
But it can be shown that current continuity is maintained once the dissipative
contribution to the current is considered\cite{BCG05}.  The equation of motion for the time
dependent density when propagating the system under the Master
Equation is given by
\ben
\label{e:masterprop}
{d\langle{n(r)}\rangle\over{dt}}|_{t=0}=-\bigtriangledown \langle j(r)\rangle +Tr(n(r)C[S]\bar{S})
\een
The last term in this equation is the contribution due to the
dissipative part of the Master Equation and so the total current is then given
by the standard expression using the current operator plus a dissipative part due
to the propagation of the system under the master equation:
\ben
\label{e:totalcurrent}
\langle j_{T}(r)\rangle=\langle j(r)\rangle+\langle j_D(r)\rangle
\een

The general many particle formulation must be simplified to an effective single
particle form to be of any use in practice.  The many body density matrix $S$
that satisfies the Liouville equation for a Hamiltonian $H$, must be rewritten
in an effective single-particle description such that the resulting single
particle density matrix $S\s$ is in terms of the eigenstates of $H\s$.  These
eigenstates are the equilibrium single particle eigenstates and are related to
$S\s$ via the expression $S\s=\sum_{lm}f_{lm}|l\rangle \langle m|$.  The density functional formulation which maps a system of interacting
particles into one of noninteracting particles with the same density is a
natural direction to proceed.
In analogy with the Hohenberg-Kohn theorem for ground-state
DFT and the Runge-Gross theorem for TDDFT, 
it can be proven that for a fixed electron-electron interaction, a given
$C[S]$ and an initial density matrix $S_{0}$, the potential is uniquely determined
by its time-dependent density $n(\br t)$.  A single particle form of equation
(\ref{e:mastereq}) can be recovered by
  applying perturbation theory for a weak interaction between the
  non-interacting electrons and the phonons in the bath, tracing out the
  irrelevant degrees of freedom, and adding the Hartree potential to the single
  particle Hamiltonian.  What results is a single particle form of the Master
  Equation with a Kohn-Sham
  version of $C[S]$ and a single particle form of the density matrix expressed
  in the basis of the equilibrium single-particle eigenstates indexed by
  $n,m$. If the expansion coefficients are related to the many body density
  matrix $S$ via $f_{nm}=tr[Sc^{\dagger}_{m}c_{n}]$, then the single particle
  master equation can be written in terms of the single particle eigenstates:
\bea
\label{e:singleMasters}
{df_{nm}\over{dt}}&=&-i\sum_{p}[H_{np}(t)f_{pm}-f_{np}H_{pm}(t)]\cr
&+&(\delta_{nm}-f_{nm})\sum_{p}(\Gamma_{np}+\Gamma_{mp})f_{pp}\cr
&-&f_{nm}\sum_{p}(\Gamma_{pn}+\Gamma_{pm})(1-f_{pp}).
\eea 
 In addition, the
  parameters in the dissipative part of the Kohn-Sham Master Equation (see
  equation (\ref{e:dissip})) can be in principal
  obtained from ground--state DFT linear response calculations, thereby eliminating the need
  for any empirical parameters.

\subsubsection{Master equation results}
\label{s:masterresults}
The master equation approach is currently under development, but some initial
results for test systems have been calculated\cite{GPC05,P06}.
The first model tested with the master equation approach was a simple 1-D double barrier resonant
tunneling system (DBRTS).
Well known results familiar to
the experimental community were derived for a DBRTS and show that the effect of inelastic
collisions, accounted for in the master equation formulation, is
important in understanding the behavior of these devices.

\begin{figure}[tbh]
\begin{center}
\leavevmode
\includegraphics[angle=270,width=8.3cm]{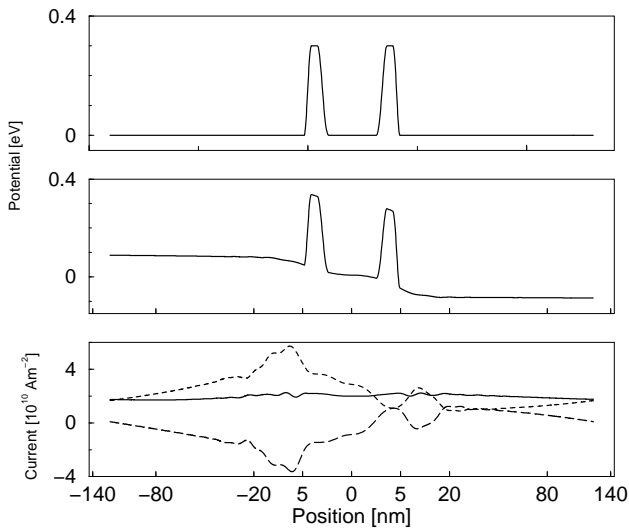}
\end{center}
\caption{Potential for the DBRTS. Top panel - no applied external field, middle panel - external field
  for the case of low dissipative coupling, $\gamma_0$. The bottom
  panel shows the total current (solid line), the Hamiltonian current
  (short dashed line) and the dissipative current (dashed line).}
\label{f:dbrts1}
\end{figure}

Among the approximations made for the phonons in this model
calculation are, that their density of states has a parabolic dependence given
by $D(\omega)\approx \omega^2$.  Furthermore, the coupling between levels $m$
and $n$ is set to a constant $\gamma_{mn}=\gamma_0$.  In the top two panels of 
Fig.~\ref{f:dbrts1}, results for the potential in the absence of a field and in
the presence of a field with low dissipative coupling $\gamma_0$ are given.  Results are similar to
the results from the Landauer formulation except for the following important
points.

The voltage drop across the device is an output of the calculation rather than
an input and there can be seen a small voltage drop across the wire as well, but
this is soon neutralized by screening effects.  

The bottom panel of Fig.~\ref{f:dbrts1} displays the total current
derived.  In keeping with the constraints of current continuity, the total
current (solid line) is constant within numerical error due to cancellation between the Hamiltonian current
(short dashed line) and the dissipative current (long dashed line).  It should
also be noted that the dissipative current is larger at the contact points,
indicating that these are the sites of local Joule heating.

\begin{figure}[tbh]
\begin{center}
\leavevmode
\includegraphics[angle=270,width=8.3cm]{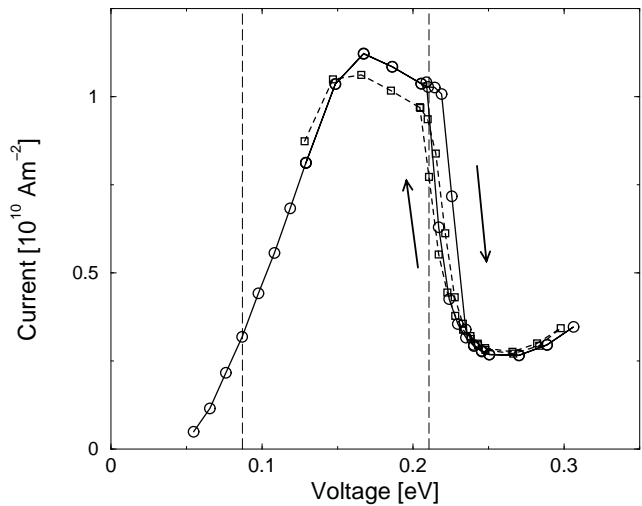}
\end{center}
\caption{Hysteresis effects for a DBRTS device as demonstrated through
  I-V plots for two different values of dissipative coupling
  $\gamma_0$.  Lower $\gamma_0$ is associated with a more pronounced hysteresis.}
\label{f:dbrts2}
\end{figure}

The Master equation approach also predicts the hysteresis effects of a DBRTS as
demonstrated in Fig.~\ref{f:dbrts2}.  At low dissipative coupling $\gamma_0$, the hysteresis is
much more evident than at higher values of $\gamma_0$. For stronger
dissipative coupling, the resonance peak also shifts towards lower voltages.

\begin{figure}[tbh]
\begin{center}
\leavevmode
\includegraphics[angle=270,width=8.3cm]{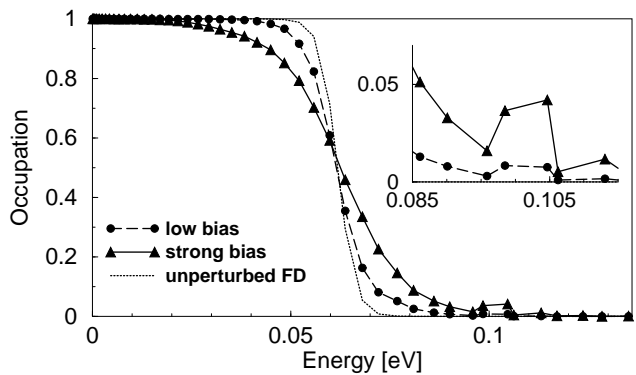}
\end{center}
\caption{Electron occupation of the resonant level in a DBRTS as given by the diagonal elements of the density matrix in the steady state.  A higher bias leads to nonlinear effects as evidenced by stronger deviations from the unperturbed Fermi-Dirac distribution.}
\label{f:dbrts3}
\end{figure}

In Fig.~\ref{f:dbrts3}, the electron
occupation of the level is plotted for different values of the applied
bias.  At higher bias, the
distribution deviates from equilibrium and exhibits a bump in the tail that
points to charging of the resonant level.  This charging is the origin of the
bistability and the hysteresis observed in DBRTs.
The bistability arises from the non-linearity associated with the
Hartree potential.
The finite current for voltages above the resonance peak stems from
the inclusion of dissipation.
In the absence of dissipative effects the current would go down to
zero the moment the resonant level becomes fully occupied and the
broadened level no longer overlaps with the Fermi level of the lead.
When dissipation is included, electrons can relax
into the leads leading to a finite current for bias voltages above the resonance.
The dissipative coupling also  controls the
size of the hysteresis effect.

\subsubsection{Chains of gold atoms}
\begin{figure}[tbh]
\begin{center}
\leavevmode
\includegraphics[angle=0,width=6.3cm]{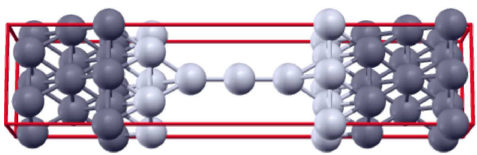}
\end{center}
\caption{Three atom gold chain: supercell geometry showing the atomic
  wire connected to two gold electrodes (Au111 surfaces).  
The dark atoms indicate the region where dissipative coupling is
present. Periodic boundary conditions are applied in all directions. 
The lateral interaction between a wire and the nearest periodic
images has negligible effect on the current.}

\label{f:goldchain}
\end{figure}
 Further calculations within the Master Equation approach 
have studied  electronic transport through a 3-atom gold wire
sandwiched between two Au(111) surfaces\cite{P06}. 
Fig.~\ref{f:goldchain}
shows the unit cell of the periodic system used in the calculations. Four layers of gold
atoms per side are included as the contacts. Dissipation is
applied in the three outermost atomic layers only.
Again, the phonon  density is assumed to be parabolic and the coupling takes the form of
 $\gamma_{ij}=\gamma\langle i|V|j\rangle$ where $\gamma$ sets the strength of
 the dissipation.

\begin{figure}[tbh]
\begin{center}
\leavevmode
\includegraphics[angle=-90,width=7.3cm]{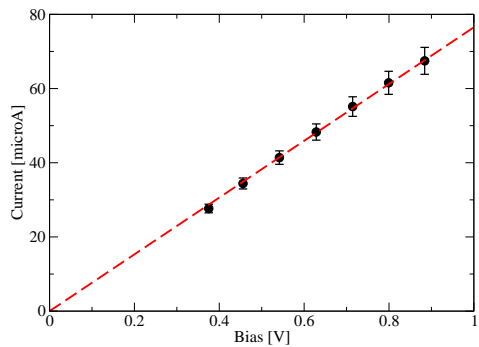}
\end{center}
\caption{ Calculated I-V characteristics of a 3-atom gold wire. The black dots
are the Master Equation results. The dashed line
indicates the characteristics corresponding to one quantum of 
conductance. The error bars reflect the fluctuations of the 
Hamiltonian current in the supercell which measure 
deviation from current continuity in the numerical calculation.}
\label{f:auiv}
\end{figure}

In this application, where a very small periodically separated cell
was used, a very large dissipative coupling was necessary to force a
steady state (this was obtained by imposing one quantum of conductance 
at once specific value of the applied bias).
 This leads to an unphysically large dissipative
current. The strength of the dissipative coupling can be reduced with
increasing system size, eventually reaching its physical value for
cells that are large compared to the electron-phonon mean free path. 
Under those conditions the Hamiltonian current dominates. It was found
that a more physical behavior of the I--V characteristics is obtained,
when only the Hamiltonian current is used to model the physical
current. Neglect of the
dissipative current in this system is further supported by the fact
that the current continuity violation of the Hamiltonian current is
weak (see Fig.~\ref{f:auiv}).

\begin{figure}[tbh]
\begin{center}
\leavevmode
\includegraphics[angle=-90,width=8.3cm]{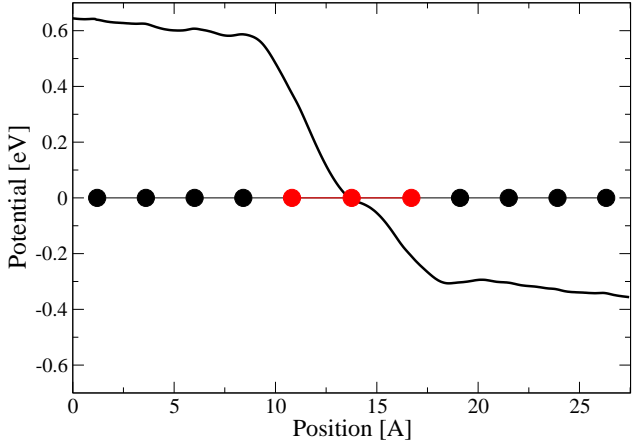}
\end{center}
\caption{3-atom gold chain --- potential:  
Total potential (including external potential, and induced Hartree and exchange-correlation potential) averaged over planes perpendicular to the wire. The external potential in the supercell is given, for illustrative purposes, in the position gauge (which does not satisfy the periodic boundary conditions). The black dots indicate the position of the atomic planes in the slab, whereas the red dots indicate the atoms of the wire. The total potential is essentially flat in the electrodes. The large drop across the wire is due to the contact resistance.}
\label{f:goldpot}
\end{figure}

 The calculated IV-characteristics, shown in Fig.~\ref{f:auiv}
 reproduces well the linear behavior measured experimentally\cite{SACY98,YRBA98,ALR03}.
 In Fig~\ref{f:goldpot} we plot the total (external plus induced)
 potential across the
 device. The potential drops occurs mainly across the length of the Au
 wire. Only a small portion of the potential drop occurs
 inside the leads. We would expect no potential drop at all inside a
 perfect metal, the small observed drop is due to the dissipation
 in the three outermost atomic layers of the leads.

\subsection{TDDFT-NEGF}
\label{s:TDDFTNEGF}

Another method that avoids the use of two external
chemical potentials with an artificial partitioning and manages to 
obtain a steady current in transport calculations is the exact non-equilibrium Green's function
approach using time-dependent density functional theory (TDDFT)~\cite{SA04,SAb04,KSAR05,SKGR06,SA06}.
The system begins in thermodynamic equilibrium in its ground-state and the leads
and device are coupled.  
A time-dependent perturbation is imposed deep inside the leads, such
that the potential exhibits a step somewhere inside the molecule.

The method applies the NEGF Keldysh formulation to the time-dependent KS equations, i.e., to a set
of non-interacting particles yielding the correct density.  
The time evolution of the system is then described by the KS equations of TDDFT 
\ben
i \dot{\Psi}\s(\br,t) = H\s(\br,t)\Psi\s(\br,t),
\een
where $H\s(\br,t)$ is the time-dependent KS-Hamiltonian, and $\Psi\s(\br,t)$ is the KS wavefunction. 
Using the continuity equation $\partial/ \partial t \,n(\br,t) = - \nabla {\mathbf j}\s(\br,t)$ for the KS current density ${\mathbf j}\s (\br,t)$
 this produces the
correct time-dependent current through the device:
\ben
I(t)=-\int{d^3r}{d\over{dt}}n(\br,t)
\een
where the integral is over some cross-section through the molecule.
In NEGF, the density can be expressed in terms of the lesser Green's function:
\ben
n(r,t)=-2iG^<(r,t;r,t)
\een
which in turn can be calculated from equations of motion for the non-interacting system.
To solve the non-interacting problem, the system is partitioned into the molecule(C), and left/right leads (L,R) as in the {\em standard approach} (see Sec.~\ref{s:DFT-NEGF}). The KS Hamiltonian can then be written as a 3x3 block matrix, yielding for the time evolution

\ben
i \frac{\partial}{\partial t} \left[
\begin{array}{c} \Psi_L\\ \Psi_C\\ \Psi_R \end{array}\right]
= \left[ \begin{array}{ccc}
H_{LL} & H_{LC} & 0\\
H_{CL} & H_{CC} & H_{CR}\\
0 & H_{RC} & H_{RR}
\end{array}\right] 
\left[ \begin{array}{c} \Psi_L\\ \Psi_C\\ \Psi_R \end{array}\right],
\een

where $\Psi_{\alpha}(\br,t)$ is the KS wavefunction projected onto left/right lead (L,R) and the molecule region(C), respectively. 

For the left and right lead ($\alpha=L,R$), we can, using the lead Greens functions $\bg_{\alpha}$ (see description in Sec. \ref{s:DFT-NEGF}), obtain an explicit solution for the projected wavefunctions:
\ben
\Psi_{\alpha}(t) = i \bg_{\alpha}(t,0)\Psi_{\alpha}(0) +\int_0^t dt' \bg_{\alpha}(t,t')H_{\alpha C}\Psi_C(t').
\een
 $\bg_{\alpha}$ is defined via $(i d/d t -H_{\alpha \alpha} (t)) \bg_{\alpha}(t,t')=\delta(t-t')$ with the appropriate boundary conditions $\bg_{\alpha}(t^{\dagger},t)=-i$ and $\bg_{\alpha}(t,t^{\dagger})=0$.

Using this, we can rewrite the expression for the molecule region as
\bea
i \frac{\partial}{\partial t} \Psi_C(t) & = &H_{CC}(t) \Psi_C(t) + \int_0^t dt' \Sigma(t,t')\Psi_C(t')\nonumber \\
 & & + i\sum_{\alpha=L,R} H_{C\alpha}\bg_{\alpha}(t,0)\Psi_{\alpha}(0).
\label{Psic}
\eea
where 
\bea
\Sigma(t,t') & = & H_{CL}(t)\, g_L (t,t')\, H_{LC}(t') \\
& & +H_{CR}(t)\, g_R (t,t')\, H_{RC}(t')\nonumber
\eea
is the self energy accounting for coupling to the leads as
 described in Sec.~\ref{s:DFT-NEGF}.
Thus Eq. (\ref{Psic}) yields, in principle exactly, the time-evolution of the molecular
wavefunction in the presence of a current through the leads.  It is non-Hermitian,
as it describes electrons flowing from left to right.
 The solution for the wavefunction of the central region is obtained by propagating an initial state, i.e.\ the ground-state of the extended sytem in equilibrium, which is obtained in a fashion analogous to the {\em standard approach}. For the actual propagation, transparent boundary conditions are imposed at the lead interfaces and a generalized Cayley method is used (see \cite{KSAR05} for details). 

The feasibility of the scheme has been tested on simple 1d systems. 
These calculations
demonstrate the independence of the steady current on the history, and show a variety
of features, such as non-monotonic dependence of current on bias, and larger transient
currents than steady-state currents.  
First applications of this approach include: (i) the study of the role of bound-states in transport\cite{S07}. Here oscillations in the density and the TDDFT KS potential have been observed, the system does not evolve towards a steady state. (ii) The coupling to nucleii\cite{VSA06}.
However, these are all non-interacting problems, and so
have not tested the procedure when interaction plays a roll.

An important issue of the formalism is whether or not a steady current can arise in the absence of any
dissipation.  For non-interacting electrons (e.g.\ the electrons in
the KS system), it has been found a steady current
develops if
\begin{enumerate}

\item
the single-particle Hamiltonian becomes time-independent as $t\to\infty$

\item
the electrodes form a continuum of states, i.e., are infinite,

\item
local density of states on the molecule is smooth.

\end{enumerate}
These all appear reasonable conditions.  Furthermore, the steady current is independent
of the history of the turning-on of the potential step, if the $t\to\infty$ Hamiltonian is.

The crucial point for the steady current is the second one.  In the presence of a continuum
of states, even non-interacting electrons dephase and a steady current develops.  This is
the mechanism for achieving a steady current in this approach, and makes dissipation to
phonons or many-body scattering effects unnecessary.  A continuum of states requires
infinite electrodes, but these are then implicitly included in source and
sink terms in the resulting equations.


\subsection{Master equation versus TDDFT-NEGF}
\label{s:comp}

Both the Master equation approach and the TDDFT-NEGF approach go beyond the
{\em standard approach}.  Each begins from a situation for which
we have
a basic theorem proving a functional-dependence of the potential on the density,
and from which the Landauer formula can be derived, at least in the case of
non-interacting electrons.   They clearly yield different results in cases
where their conditions differ, such as when dissipation is strong in the Master
equation, but it is as yet unknown if they differ when applied to the same
situation (and if they do, which one is `correct').
In this section, we compare and contrast the two different
approaches.

{\em Basic variable:}  In the Master equation approach, because the system experiences
a solenoidal magnetic field and periodic boundary conditions are used, the basic
variable is the current density.  In contrast, the TDDFT-NEGF approach has been developed
using the density itself as the basic variable.  While this is more familiar within DFT,
the current allows development of simple approximations such as Vignale-Kohn, 
yielding
simple corrections to the conductance.

{\em Boundary conditions:}  Almost all present calculations are performed on localized
systems embedded between two electrodes, using localized basis sets, and this is also
true for TDDFT-NEGF.  The Master equation approach both requires and allows use of plane-wave
codes, and so can make it much easier to adapt present solid-state codes for use in
transport calculations.

{\em Fundamental theorems:}  TDDFT-NEGF is based on the Runge-Gross theorem, which applies
only to finite systems.  Yet the leads must be infinite to produce the required dephasing.
This is an inconsistency in the approach whose implication is
debatable. The Master equation, on the other hand,
required proving the Runge-Gross theorem for the Master equation instead of the time-dependent
Schr\"odinger equation, and so requires introducing new functionals.  These may (or may not)
reduce to the standard ones of TDCDFT in the limit of weak dissipation.

{\em Need for dissipation:}  The TDDFT-NEGF approach demonstrates that a steady current can
arise {\em without} explicit dissipation mechanisms, once the leads are infinite.  The Master
equation approach may well reduce to the same result in the limit in which the ring size is
very large and the dissipation small, but this has yet to be demonstrated.  If so, they become,
in that limit, simply two different procedures for finding the same result.  If not, one might
well be correct for molecular conductance, the other not.  

{\em Weak bias:}  With small dissipation,
the Master equation produces the same conductance in the zero-bias limit
as the Kubo response \cite{BCG05}, and therefore includes the XC corrections to the Landauer
formula as seen in section\ref{s:dcTrans}.  
A similar result can be derived from the TDDFT-NEGF formula\cite{SKGR06},
although couched
in DFT terms.  Thus the formalisms agree in the limit of weak bias and small dissipation,
even for interacting electrons.
For the description of Joule heating effects and phonon scattering,
which are not easily incorporated into the TDDFT-NEGF approach,
the Master Equation formalism provides a natural framework.

\section{Summary}
\label{s:summ}

In this brief, non-comprehensive review, we have critically examined the
present state-of-the-art of DFT calculations of transport through single
molecules.   Our findings are:

\begin{itemize}

\item
Even the steady state of current flowing through a molecule is not
included by the basic theorems establishing ground-state DFT.

\item
The commonly used approximation of {\em ground-state} DFT in the
Landauer formula, which we dub the {\em standard approach}, has a
 variety of limitations, making it inexact, even if the exact ground--state functional were known
and used.

\item
Standard density functional approximations, such as LDA, GGA, or hybrids,
are insufficiently accurate to treat molecules weakly coupled to leads,
and likely produce large overestimates of the current.  This effect
might be the origin of the overestimates relative to experiment.
Orbital-dependent functionals, such as exact-exchange or self-interaction
corrected LDA (LDA-SIC), should perform much better.

\item
The {\em standard approach} is only a Hartree-level theory for the conductance, and neglects
non-local XC corrections to the conductance.  This is demonstrable in the
case of weak bias.  Either orbital-dependent or current-dependent functionals
are needed to even estimate these corrections.

\item
For finite bias, several approaches have been developed that are within
time-dependent DFT frameworks, thus addressing
the problem of the inadequate ground-state approximation. Two of these
have been described in this review. (i) The Master equation approach 
which includes dissipation to phonons. 
(ii) The TDDFT-NEGF formalism whis does not have the need for dissipation.
 Connections are being developed between the two, and
time will tell which is more practical, reliable, or relevant.

\end{itemize}

Open questions for both new approaches and any other DFT treatments
include the following:

\begin{itemize}
\item
Do they agree with the Kubo response weak-bias limit discussed in Sec.~\ref{s:dcTrans}?
Both new formalisms do this.

\item
At finite bias, which effects are included or not in each approach?

\item
At finite bias, in what limits do they agree or disagree with
each other and with the standard approach?  We already know that
within ALDA, all give the same answer as the standard approach,
but expect differences with non-local non-adiabatic functionals.

\item
Does one always approach a steady state, and is it
unique?  The hysterises
seen in model calculations with the Master equation
is an example of more than one steady solution.

\item
Is there a dependence on how the potential is turned on?
Recently, bound states of the molecule have been shown to
lead to infinitely oscillating contributions to the
current in model calculations using the TDDFT-NEGF
approach, when the turn-on is non-adiabatic.  Do these
survive in an interacting system?

\item
Are there infinite memory effects in the time-dependent
Kohn-Sham potential?  These are logically possible,
and might even be necessary, to reproduce the physics.

\item
Exactly what features of these theories are needed to
reproduce strongly-correlated effects such as Coulomb
blockade, and is there any chance to model such features?

\end{itemize}

On the other hand,  the present
Landauer-type calculations (the standard approach) yield
the correct steady solution to the more sophisticated approaches
{\em when the functional is local in time and space} (see Sec \ref{s:dcTrans}), and
this may be sufficient for many purposes.
Only more demanding calculations (be they time-dependent DFT, non-local
static DFT, or CI or GW) and
better characterized experiments can tell us what is important to
reliable first-principles predictions of the conductance of single
molecules.

The authors are very  grateful to 
H.\ Baranger, F.\ Evers, R.\ Gebauer, N.\ Lang, S.\ Piccinin, E.\ Prodan, S.\ Sanvito, and W.\ Yang for many fruitful discussions and their contributions. 
MK acknowledges support by the
Deutsche Forschungsgemeinschaft (grant KO3459/1). This work was
supported by the DOE under grant DE-FG02-01ER45928.

\end{document}